\documentclass[pdflatex,sn-nature]{sn-jnl}

\usepackage{siunitx}
\usepackage{graphicx}%
\usepackage{multirow}%
\usepackage{amsmath,amssymb,amsfonts}%
\usepackage{amsthm}%
\usepackage{mathrsfs}%
\usepackage[title]{appendix}%
\usepackage{xcolor}%
\usepackage{textcomp}%
\usepackage{manyfoot}%
\usepackage{booktabs}%
\usepackage{algorithm}%
\usepackage{algorithmicx}%
\usepackage{algpseudocode}%
\usepackage{listings}%
\usepackage{mathrsfs}
\usepackage{float}
\usepackage{type1cm}
\usepackage{fix-cm}

\DeclareSIUnit\angstrom{\protect \text {Å}}

\theoremstyle{thmstyleone}%
\theoremstyle{thmstyletwo}%

\theoremstyle{thmstylethree}%
\usepackage{xspace}                                   
\usepackage[version=4]{mhchem}

\newcommand{\NPS}{\ce{Na3PS4}\xspace}          
\newcommand{\htNPS}{t-NPS\xspace}
\newcommand{\bmNPS}{c-NPS\xspace}

\newcommand{\Q}{\ensuremath{\mathbf{Q}}\xspace}          
\newcommand{\Qmag}{\ensuremath{|\mathbf{Q}|}\xspace}
\newcommand{\Qnot}{\ensuremath{|\mathbf{Q}_{0}|}\xspace}
\newcommand{\corrLength}{\ensuremath{\xi}\xspace}
\newcommand{\spacing}{\ensuremath{d}\xspace}
\newcommand{\spacingEqn}{\ensuremath{2\pi}/\Qnot}

\newcommand{\TTC}{\ensuremath{C}\xspace}
\newcommand{\TTCFull}{\ensuremath{C(\mathit{\mathbf{Q}},t_1,t_2)}\xspace}
\newcommand{\ISF}{\ensuremath{|F(\textit{\textbf{Q}},\tau)|}\xspace}
\newcommand{\ISFIP}{\ensuremath{|F(\textit{\textbf{Q}},\tau)|^2}\xspace}

\newcommand{\gTwoFull}{\ensuremath{g_{2}(\textit{\textbf{Q}},\tau)}\xspace}
\newcommand{\gTwo}{\ensuremath{g_{2}}\xspace}
\newcommand{\tone}{\ensuremath{t_1}\xspace}
\newcommand{\tage}{\ensuremath{t_{age}}\xspace}

\newcommand{\shape}{\ensuremath{\beta}\xspace}
\newcommand{\taur}{\ensuremath{\tau_{r}}\xspace}

\newcommand{\conductivity}{\ensuremath{\sigma}\xspace}

\newcommand{\deviatoric}{\ensuremath{\epsilon'(p)}\xspace}
\newcommand{\strain}{\ensuremath{\epsilon(\mathbf{r})}\xspace}
\newcommand{\meanstrain}{\ensuremath{\langle\epsilon\rangle}}

\newcommand{\eigenstrain}{\ensuremath{\epsilon^0(p)}\xspace}

\newcommand{\Tpath}[3]{\qty{#1}{\kelvin} $\rightarrow$ \qty{#2}{\kelvin} $\rightarrow$ \qty{#3}{\kelvin}\xspace}

\newcommand{\Tr}[1]{\operatorname{Tr}(#1)}

\begin{document}
\title[Article Title]{Strain-Glass Dynamics in the Superionic Conductor \NPS}


\author*[1,2]{\fnm{Erik S.} \sur{Lamb}}\email{elamb@ucsd.edu}

\author[1,2]{\fnm{Nicol\`{o}} \sur{D'Anna}}

\author[3]{\fnm{Phillip} \sur{Ridley}}

\author[4,5]{\fnm{Nelson} \sur{Hua}}

\author[6]{\fnm{Ivan} \sur{Zaluzhnyy}}

\author[2,7]{\fnm{James} \sur{Burrow}}

\author[1,2]{\fnm{Emilio} \sur{Escauriza}}

\author[1]{\fnm{Robin} \sur{Glefke}}
\author[1]{\fnm{Boyan K.} \sur{Stoychev}}
\author[1,2]{\fnm{Mingze} \sur{Wan}}


\author[8]{\fnm{Zhe} \sur{Ren}}
\author[8]{\fnm{Fabian} \sur{Westermeier}}
\author[8]{\fnm{Michael} \sur{Sprung}}

\author[1]{\fnm{Alex} \sur{Fra\~{n}\'{o}}}

\author[2,3,7]{\fnm{Ying Shirley} \sur{Meng}}\email{shirleymeng@uchicago.edu}
\author*[1,2]{\fnm{Oleg G.} \sur{Shpyrko}}\email{oshpyrko@physics.ucsd.edu}

\affil*[1]{\orgdiv{Department of Physics}, \orgname{University of California San Diego}, \orgaddress{\street{9500 Gilman Dr}, \city{La Jolla}, \state{California}, \postcode{92093}, \country{United States of America}}}

\affil[2]{\orgdiv{Energy Storage Research Alliance}, \orgname{Argonne National Laboratory}, \orgaddress{\street{9700 South Cass Avenue}, \city{Lemont}, \state{Illinois}, \postcode{60439}, \country{United States of America}}}

\affil[3]{\orgdiv{Department of NanoEngineering}, \orgname{University of California San Diego}, \orgaddress{\street{9500 Gilman Dr}, \city{La Jolla}, \state{California}, \postcode{92093}, \country{United States of America}}}

\affil[4]{\orgdiv{PSI Center for Photon Science}, \orgname{Paul Scherrer Institute}, \orgaddress{\street{Forschungsstrasse 111}, \postcode{5232} \city{Villigen PSI}, \country{Switzerland}}}

\affil[5]{\orgdiv{Institute for Quantum Electronics}, \orgname{ETH Zürich}, \orgaddress{\postcode{8093} \city{Zürich}, \country{Switzerland}}}

\affil[6]{\orgdiv{Institute of Applied Physics}, \orgname{University of Tübingen}, \orgaddress{\street{Auf der Morgenstelle 10}, \postcode{72076} \city{Tübingen}, \country{Germany}}}

\affil[7]{\orgdiv{Pritzker School of Molecular Engineering}, \orgname{The University of Chicago}, \orgaddress{\street{5801 S Ellis Ave}, \city{Chicago}, \state{Illinois}, \postcode{60637}, \country{United States of America}}}

\affil[8]{\orgname{Deutsches Elektronen-Synchrotron DESY}, \orgaddress{\street{Notkestr. 85}, \postcode{22607} \city{Hamburg}, \country{Germany}}}

\abstract{Ionic conduction in solid-state electrolytes is conventionally described as cations migrating through a static host lattice. Yet the fastest conductors are often the most disordered, and how disorder shapes mesoscale transport remains unresolved. Using Bragg X-ray photon correlation spectroscopy on a micrometer-sized ball-milled \NPS particle, we find that the host is not static. The cubic phase behaves as a strain glass of three degenerate tetragonal variants whose domain walls move under residual milling stress, crossing over irreversibly to thermally activated dynamics on annealing. Rearrangements occur as scale-free avalanches, as in ferroelastic martensites but without external driving, while the tetragonal polymorph shows none of these features and relaxes fifty times more slowly. These dynamics evolve on timescales comparable to ionic diffusion across the particle. This work introduces a probe of dynamics in locally frustrated solid electrolytes, revealing that mechanochemical synthesis yields not merely a metastable polymorph, but a stress-driven strain glass.}

\maketitle


Despite rapid advances, the local mechanisms leading to anomalous superionic transport in solid-state electrolytes (SSEs) remain poorly understood~\cite{famprikis_fundamentals_2019}. While the chemical composition and structural symmetries of SSEs are diverse, making a universal explanation elusive, numerous superionic conductors exhibit fundamentally frustrated degrees of freedom, whether chemical, compositional, or structural~\cite{tsukasaki_direct_2017,breuer_rapid_2018}. Such frustration flattens the energy landscape ions explore, increasing the probability of statistically rare ion hopping events by orders of magnitude~\cite{wood_paradigms_2021}. Disorder is therefore often a better descriptor of functionality in these materials than idealized crystalline order. This recognition is playing a central role in SSE development, where defect engineering and the synthesis of amorphous or metastable phases have improved ionic conductivity, a prerequisite for the widespread adoption of all-solid-state batteries (ASSBs)~\cite{maus_discrepancy_2023, landgraf_disorder-mediated_2025}.

How ion motion couples to the dynamics of the host lattice itself has been examined almost entirely at the picosecond-to-nanosecond timescales accessible to atomistic simulation and ultrafast probes~\cite{poletayev_defect-driven_2022, poletayev_persistence_2024}, leaving the slow, collective structural dynamics of disordered landscapes uncharacterized. In the sodium thiophosphate \NPS, nuclear magnetic resonance and molecular dynamics indicate that bulk \ce{Na+} mobility is nearly identical in the cubic and tetragonal polymorphs, implying that the conductivity difference arises not from intrinsic \ce{Na+} mobility but from higher defect density and local strain of the metastable cubic phase~\cite{krauskopf_local_2018}, distributed on nano- to mesoscale lengths that ensemble-averaged probes cannot resolve. Therefore, whether and how ionic migration at the nano- to mesoscale is modulated by local disorder and symmetry variations via induced strain, rather than simple site occupancy variations, remains an open question.

Sodium thiophosphates are a promising class of sodium-ion SSEs, exhibiting the forms of frustration and structural polymorphism described above, fast ionic conductivity, and intrinsically soft mechanical properties~\cite{hayashi_superionic_2012,hayashi_high_2014,krauskopf_local_2018}. \NPS crystallizes into two room-temperature polymorphs depending on synthesis protocol. Jansen et al. first demonstrated synthesis of tetragonal \htNPS by heat treatment to \qty{1023}{\kelvin} followed by slow annealing, with a room-temperature ionic conductivity of \mbox{\conductivity = \qty{4.17e-6}{\siemens\per\centi\meter}~\cite{jansen_synthesis_1992}.} Hayashi et al. later stabilized a cubic phase, \bmNPS, at room temperature by mechanochemical milling, with \conductivity = \qty{4.6e-4}{\siemens\per\centi\meter}~\cite{hayashi_superionic_2012,hayashi_high_2014}. Subsequent syntheses of the tetragonal phase reached \mbox{$\sim$\qty{e-5}{\siemens\per\centi\meter}~\cite{famprikis_under_2020}}, and the cubic phase is consistently found to be about an order of magnitude more conductive, despite a deviation in lattice parameters of only $\sim$\qty{0.2}{\percent}. Krauskopf et al. showed that while \bmNPS is globally cubic, its local order is better described as tetragonal, attributing the enhanced conduction to increased local strain, disorder, and defect density in the metastable phase~\cite{krauskopf_local_2018}. A cubic parent structure with three energetically degenerate, local tetragonal distortions invites comparison with ferroelastic materials, in which the strain tensor plays the role of an order parameter. Ferroelastics and martensites are characterized by hysteresis, memory effects, and intermittent avalanche dynamics with scale-free statistics~\cite{salje_ferroelastic_2012,porta_intermittent_2018}. Memory in fast-ion conductors has been reported at the level of correlated single-site hopping~\cite{poletayev_persistence_2024}. However, the collective, configurational memory and scale-free rearrangement statistics reported here have not, to our knowledge, been observed in an SSE.

Here we report measurements of nanoscale structural dynamics in a solid electrolyte using Bragg X-ray photon correlation spectroscopy (XPCS). Tracking a single \bmNPS primary particle through the annealing series \Tpath{298}{405}{350}, we find signatures of athermal, stress-driven domain wall motion with a narrow distribution of timescales at room temperature, crossing over monotonically to thermally activated dynamics on annealing as the residual stress from milling is depleted~\cite{frey_liquid-like_2025}. The relaxation is strongly non-Arrhenius and irreversible. Discrete rearrangement events follow scale-free statistics with an exponent coincident with values measured for microstructural avalanches in ferroelastic martensites, suggesting similar mechanisms and structural symmetry, but occurring here under isothermal conditions with no external drive. A tetragonal reference measured under identical conditions shows dynamics more than fifty times slower and possesses none of these complex features. The progression from stress-driven collective motion toward uncorrelated diffusive dynamics parallels the loss of ionic conductivity reported for annealing into the tetragonal ground state~\cite{hayashi_high_2014,famprikis_fundamentals_2019}, suggesting that the local stress state of the host lattice, and not composition alone, is a variable available for tuning transport in mechanochemically synthesized SSEs. More broadly, these results establish Bragg XPCS as a direct probe of the slow, collective host dynamics in locally frustrated solid electrolytes, a regime inaccessible to atomistic simulation, ultrafast probes, and ensemble-averaged measurements.

\begin{figure}[ht!]
    \centering
    \includegraphics[width=1\linewidth]{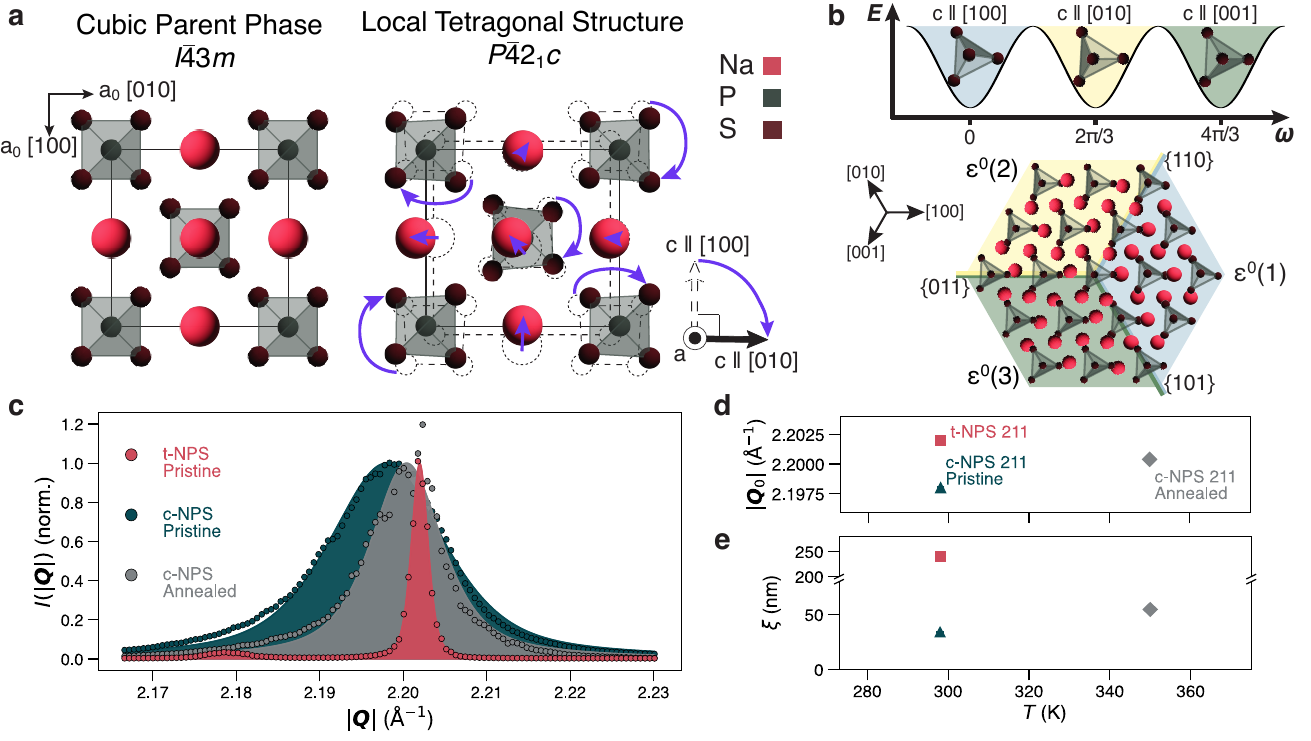}
\caption{\textbf{Structural Motifs of \NPS.} \textbf{a} Crystal structure showing the cubic parent phase and the local tetragonal distortion. Three degenerate, tetragonal distortions along [100], [010], and [001] are related by the eigenstrain \eigenstrain. Re-orientations of \eigenstrain are not a rigid body rotation of the unit cell but occur via collective internal anion and cation rearrangements (violet arrows)~\cite{khachaturyan_2008}. \textbf{b} $E(\omega)$ schematically represents the energy of the three degenerate deviatoric strain variants (Methods) as a function of angle in order-parameter space. Habit planes \{110\}, \{011\}, and \{101\} of the three degenerate tetragonal variants, where two variants coexist. The core where all three variants coexist is a disclination. Such defects are unstable in isolation and form pairs of opposite handedness that can annihilate~\cite{kleman_disclinations_2008}. \textbf{c} Azimuthally integrated intensity of the 211 Bragg peak for pristine (\qty{298}{\kelvin}), annealed (\qty{350}{\kelvin}) \bmNPS, and reference \htNPS, normalized to the fitted peak amplitude. Solid profiles are pseudo-Voigt fits. The tetragonal phase shows splitting into the lower-\Qmag 112 reflection. \textbf{d,e} Peak position \Qnot and correlation length \corrLength from the fits in \textbf{c}. On annealing, \Qnot shifts from \qty{2.198}{\per\angstrom} toward the tetragonal 211 position at \qty{2.202}{\per\angstrom}, and \corrLength increases by \qty{60}{\percent} from \num{34.2}$\pm$\num{0.5}~\unit{\nano\meter} to \num{54.7}$\pm$\num{1.7}~\unit{\nano\meter}, consistent with progressive ordering of the variant populations. The tetragonal reference has \corrLength = \num{243.2}$\pm$\num{2.7}\unit{\nano\meter}.
    }\label{fig:fig_intro}
\end{figure}

\section*{Results}\label{sec_results}
\noindent\textbf{Strain Glass}

Ball-milled \bmNPS was first described as a ``glass ceramic,'' with its order-of-magnitude higher conductivity attributed to a mechanochemically stabilized cubic phase~\cite{hayashi_superionic_2012,hayashi_high_2014}. Subsequent work established instead that it is a single crystalline phase whose long-range average is cubic but whose local structure retains the tetragonal motifs of the \htNPS ground state~\cite{jansen_synthesis_1992,krauskopf_local_2018,famprikis_under_2020} (Fig.~\ref{fig:fig_intro}\textbf{a,b}). Although the $c/a$ lattice parameter ratio was shown to deviate only slightly from unity~\cite{krauskopf_local_2018}, the local tetragonality has been confirmed independently by reflections forbidden in the cubic structure, pair-distribution-function analysis~\cite{famprikis_under_2020}, and splitting of the \ce{PS4^3-} stretching and bending modes in Raman spectra~\cite{nguyen_single-step_2019,famprikis_under_2020}.

As in ferroelectrics, magnets, and ferroelastic martensites, where a high-symmetry parent distorts into degenerate lower-symmetry variants~\cite{mangu_hidden_2025,shpyrko_direct_2007,artemev_three-dimensional_2001}, the cubic-to-tetragonal distortion in \bmNPS takes the form of three degenerate strain variants \eigenstrain, $p=(1,2,3)$, each describing the same small tetragonal distortion along a different cubic axis (Fig.~\ref{fig:fig_intro}\textbf{b} and Methods). Each variant can be decomposed into a dilatational component, which changes the unit cell volume, and a deviatoric component, which changes its shape. The deviatoric contribution thus carries the symmetry-breaking tetragonal distortion and the elastic interaction between variants, and exceeds the dilatational contribution by an order of magnitude (Methods), making it the relevant order parameter~\cite{barsch_twin_1984}. Because the three variants are related by cubic symmetry, equal populations within a volume average to zero strain, $\langle\strain\rangle_V=0$ (Methods). The broad cubic 211 peak in Fig.~\ref{fig:fig_intro}\textbf{c} is therefore consistent with tetragonally strained domains of roughly equal population, with a correlation length \corrLength = \num{34.2}$\pm$\num{0.5}~\unit{\nano\meter} extracted from the pseudo-Voigt fit (Fig.~\ref{fig:fig_intro}\textbf{e}), in agreement with prior diffraction studies~\cite{famprikis_under_2020,krauskopf_local_2018}.

The pristine ball-milled phase has tetragonally strained order extending only over $\sim$\corrLength, with no long-range strain across the particle size of $L\sim\qty{1.32}{\micro\meter}$ (Section S4, Supplementary Materials). This places \bmNPS in a class of non-equilibrium ferroelastic systems known as \textit{strain glasses}~\cite{bossen_frozen_2024,ren_strain_2010,ren_strain_2014}. These are materials that appear cubic on average but consist of a disordered mosaic of nanoscale, locally distorted regions, because competing variants and defects frustrate long-range order. Whether this mosaic is static, or rearranges on timescales relevant to ion transport, cannot be determined from diffraction alone and requires a direct probe of its mesoscale dynamics.

\noindent\textbf{Bragg X-ray Photon Correlation Spectroscopy}
\begin{figure}[ht!]
    \centering
    \includegraphics[scale=0.7]{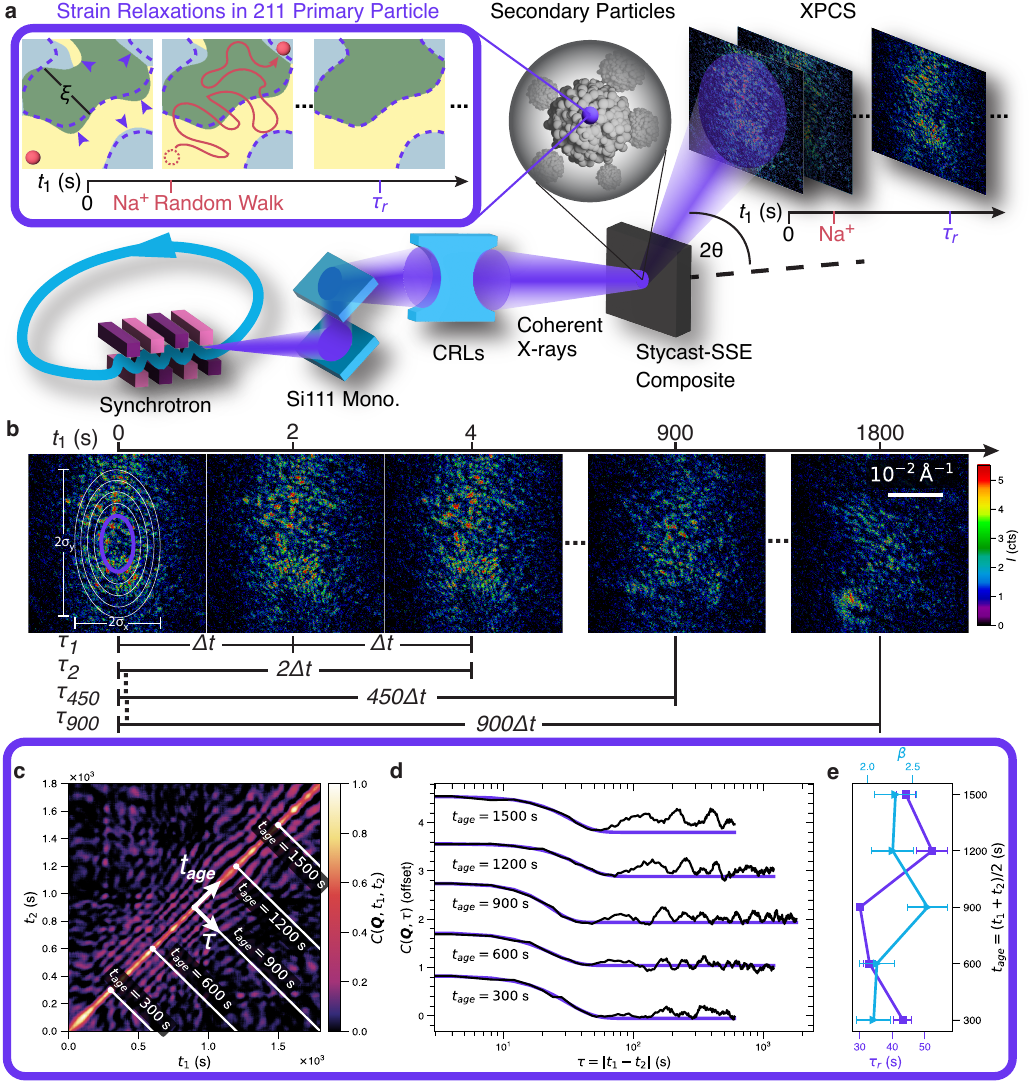}
    \caption{\textbf{Bragg X-Ray Photon Correlation Spectroscopy.} \textbf{a} Experimental schematic for single primary particle Bragg XPCS measurements showing sensitivity to collective domain wall motion of degenerate strain variants with correlation lengths on the order of \corrLength. \textbf{b} XPCS scan showing the speckle evolution of the 211 peak for pristine \bmNPS with a \qty{2}{\second} acquisition rate and total measurement time of \qty{1800}{\second} along the time axis. Correlation functions are calculated by partitioning an elliptical profile with semi-axes $1\sigma_x\times1\sigma_y$ in size, determined from a two-dimensional Gaussian fit, into five concentric elliptical rings of equal integrated intensity and correlating the speckle change within each partition. The delay time $\tau_n$ is calculated for each possible frame separation $n\Delta t$. \textbf{c,d} Two-Time Correlation \TTC matrix showing complex recorrelation events, a signature of underlying disorder in \bmNPS. Line-cuts at various \tage are shown to the right in \textbf{d}, demonstrating an alternative means of calculating \gTwo as a function of $\tau$ directly from the \TTC matrices by averaging every semi-antidiagonal line-out of the \TTC for every \tage. Consequently, short delay times have better statistics and lower uncertainty than longer delay times. \textbf{e} KWW equation~(\ref{eq:KWW}) fit results to the line-outs in \textbf{d}, showing heterogeneity with age but no exponential or power-law dependence. }
    \label{fig:fig_xpcs}
\end{figure}

When coherent X-rays scatter from a crystal, waves from different regions interfere to produce a grainy pattern known as speckle (Fig.~\ref{fig:fig_xpcs}\textbf{a}), a fingerprint of the exact nanoscale arrangement of the material. A static arrangement gives a frozen speckle pattern; as domains move, the speckle changes. X-ray photon correlation spectroscopy (XPCS) measures how quickly the speckle loses its resemblance to itself, and therefore how fast the underlying structure rearranges~\cite{sutton_observation_1991,krivoglaz_x-ray_1996}. This makes a strain glass well suited to XPCS studies at a Bragg peak. The peak position is set by the average lattice, while the speckle within it is sensitive to the local strain configuration, so the measured dynamics report directly on the motion of tetragonal domain walls, the strain-glass order parameter (Methods). The choice of reflection is consequential. For 211, the unequal \textit{hkl} Miller indices give the three variants different projections $\Q\cdot\epsilon^0(p)$ onto the momentum transfer \Q, distinguishing the variant with $c$ along [100] from the other two. A reflection with equal indices, such as 111, would project identically onto all three variants and could not distinguish them (Section S3, Supplementary Materials). We quantify these dynamics with the autocorrelation function \gTwoFull and the two-time correlation function \TTCFull, which tracks how the speckle evolves over the measurement (Fig.~\ref{fig:fig_xpcs}\textbf{b--e} and Methods).

The selected anti-diagonals extracted from \TTCFull in Fig.~\ref{fig:fig_xpcs}\textbf{d} show a similar leading relaxation followed by quasi-periodic recorrelations. For a single \taur relaxation, the absolute value of the intermediate scattering function (ISF) \ISF can be expressed as the Kohlrausch-Williams-Watts (KWW) stretched exponential~\cite{kohlrausch_theorie_1854,williams_non-symmetrical_1970}
\begin{equation}\label{eq:KWW}
    \ISF = \exp{\left[-(\tau/\taur)^{\shape}\right]}
\end{equation}
where \shape is the shape parameter that characterizes the nature of the observed dynamics. Fit results for the five $t_{age}$ anti-diagonals shown in Fig.~\ref{fig:fig_xpcs}\textbf{e} reveal a mean relaxation time $\langle\taur\rangle = \num{40.6}\pm\num{3.6}~\unit{\second}$ and a mean shape parameter of $\langle\shape\rangle = \num{2.3}\pm\num{0.1}$, indicative of driven, ballistic dynamics~\cite{frey_liquid-like_2025}. The five cuts are a sparse sampling of $t_{age}$, and the scatter is consistent with fit uncertainty rather than systematic drift. In glasses, typically \taur has an exponential or power-law dependence on $t_{age}$~\cite{ruta_atomic-scale_2012,danna_dynamic_2026} that we do not observe on the timescales of our thermally driven measurements, justifying our consistent use of the KWW formalism for subsequent characterization of symmetry and temperature dependent dynamics. 

More notable than fluctuations in \taur and \shape are recorrelations, following the leading decays in Fig.~\ref{fig:fig_xpcs}\textbf{d}, where the correlation partially recovers, meaning the nanoscale structure returns toward an earlier configuration. These indicate configurational memory stemming from domain walls weakly pinned at defects such as disclination pairs or cubic-phase impurities. Their stress-driven displacements are confined to a limited region of configuration space, a metabasin, rather than evolving irreversibly, as seen in disordered magnetic systems~\cite{pierce_disorder-induced_2005}. A residual analysis of \bmNPS at \qty{298}{\kelvin} (Fig. S9, Supplementary Materials) reveals an average quasi-periodic structure with period $\Delta\tau=\num{142.6}\pm\num{2.1}~\unit{\second}$. Oscillations in \TTCFull have also been associated with shear flow arising from stress release at random locations in systems such as polymer melts and colloidal gels~\cite{song_one-dimensional_2019,rogers_echoes_2014}, and are consistent with the stress-driven domain wall motion indicated by $\shape\approx2$ as well as disclination flow along the habit planes shown in Fig.~\ref{fig:fig_intro}\textbf{b}.

\noindent\textbf{Symmetry Dependent Dynamics}
\begin{figure}[ht!]
    \centering
    \includegraphics[scale=0.55]{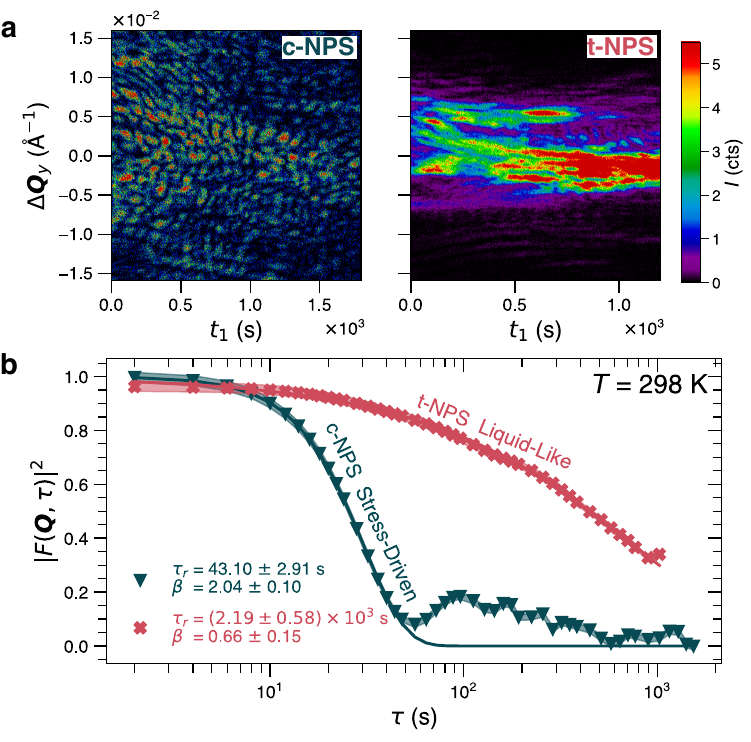}
    \caption{\textbf{Waterfalls and Intermediate Scattering Function Comparison.} \textbf{a} Waterfall plots made by taking a vertical slice with span $\Delta \Q_{y}\approx\pm\qty{1.5e-2}{\per\angstrom}$ at the center of the Bragg peak $\Delta \Q_{x}=0$ and propagating in time \tone. \textbf{b} \ISFIP showing the KWW fit results in the legend with a factor $\sim$\num{50} longer and more stretched, liquid-like dynamics ($\shape<1$) in \htNPS over compressed, stress-drive \bmNPS ($\shape\sim2$) along with long $\tau>\taur$ recorrelations, both at \qty{298}{\kelvin}.}
    \label{fig:fig_tet}
\end{figure}

Comparison of cubic and tetragonal symmetries reveals a dramatic difference in the nanoscale texture and dynamics between the two \NPS polymorphs. From the azimuthally \Q-integrated Bragg scattering intensity in Fig.~\ref{fig:fig_intro}\textbf{c}, it is immediately evident that the peak width of 211 \htNPS, at $\Delta \Q_{x}\approx 2\pi/\corrLength =$ \qty{2e-3}{\per\angstrom}, is significantly narrower than the pristine \bmNPS 211 width of $\Delta \Q_{x}\approx$ \qty{2e-2}{\per\angstrom}, both measured at \qty{298}{\kelvin}. This is corroborated by the narrow span in $\Delta\Q_y$ of approximately $\pm$\qty{0.5e-2}{\per\angstrom} for \htNPS compared to $\pm$\qty{1.5e-2}{\per\angstrom} for \bmNPS in the waterfall plot in Fig.~\ref{fig:fig_tet}\textbf{a}. The substantively less diffuse, nearly single speckle Bragg peak in ($\Delta \Q_{x}$,$\Delta \Q_{y}$) of \htNPS points to a more ordered structure with a larger \corrLength= \num{243.2}$\pm$\num{2.7}~\unit{\nano\meter} and a primary particle size larger than the coherent volume (Section S4, Supplementary Materials). This contrasts with the disordered, strain glass structure describing \bmNPS. Together with the synthesis protocol detailed in Methods, it is apparent that the more ordered structure of \htNPS is due to the longer heat-treatment and slow anneal, making the tetragonal polymorph the preferred ground-state and less defective than the quenched, metastable cubic polymorph~\cite{famprikis_fundamentals_2019,krauskopf_local_2018}. 

Correspondingly, as the less defective and less strained structure, at \qty{298}{\kelvin} \htNPS exhibits a $\taur= (2.19\pm0.58)\times10^3~\unit{\second}$, a factor of $\sim$\num{50} larger than \taur= $43.10\pm2.91~\unit{\second}$ observed for cubic phase \NPS, with $\taur$ for the tetragonal phase exceeding the measurement window and therefore representing a lower bound. Differences in the shape parameter \shape are also telling, with \htNPS showing a stretched $\shape \sim 0.7$, indicating a broad distribution of independent sub-diffusive relaxation timescales, versus compressed dynamics with $\shape\sim\num{2}$, the signature of ballistic, stress-driven relaxation rather than thermally activated diffusion~\cite{frey_liquid-like_2025}. A partially recurrent, quasi-periodic speckle pattern in the cubic phase is seen as recorrelations in \ISFIP in Fig.~\ref{fig:fig_tet}\textbf{b} for $\tau > \taur$, beyond the leading relaxation, absent from the single time-scale intermediate scattering function decay for \htNPS. This is visible in the waterfall plots in Fig.~\ref{fig:fig_tet}\textbf{a}, with recurrent, over-sampled speckle configurations persisting in \bmNPS contrasted with a smoothly varying speckle for \htNPS. 

\bmNPS exhibits diffuse scattering, a fast \taur, a highly compressed shape parameter \shape, and recorrelations at long $\tau > \taur$ arising from partial returns in speckle configurations. These are different expressions of a single underlying strain glass state characterized by short-range order and weakly pinned domain walls in a frustrated, stress-driven configuration stabilized by a high defect concentration~\cite{ren_strain_2010,ren_strain_2014}. This is particularly pronounced when compared with the slow and non-recurrent dynamics of the annealed ground state symmetry of \htNPS which, although possessing the same local symmetry as \bmNPS, has the variant degeneracy lifted by long-range tetragonal order. Therefore, dynamics measured in \bmNPS are consistent with fluctuations of the degenerate, local tetragonality that constitutes the strain glass order parameter, arising from the mechanochemical synthesis and is contrasted with the slower, annealed \htNPS. 

\noindent\textbf{Strain-Glass Dynamical Heterogeneities}
\begin{figure}[ht!]
    \centering
    \includegraphics[width=1\linewidth]{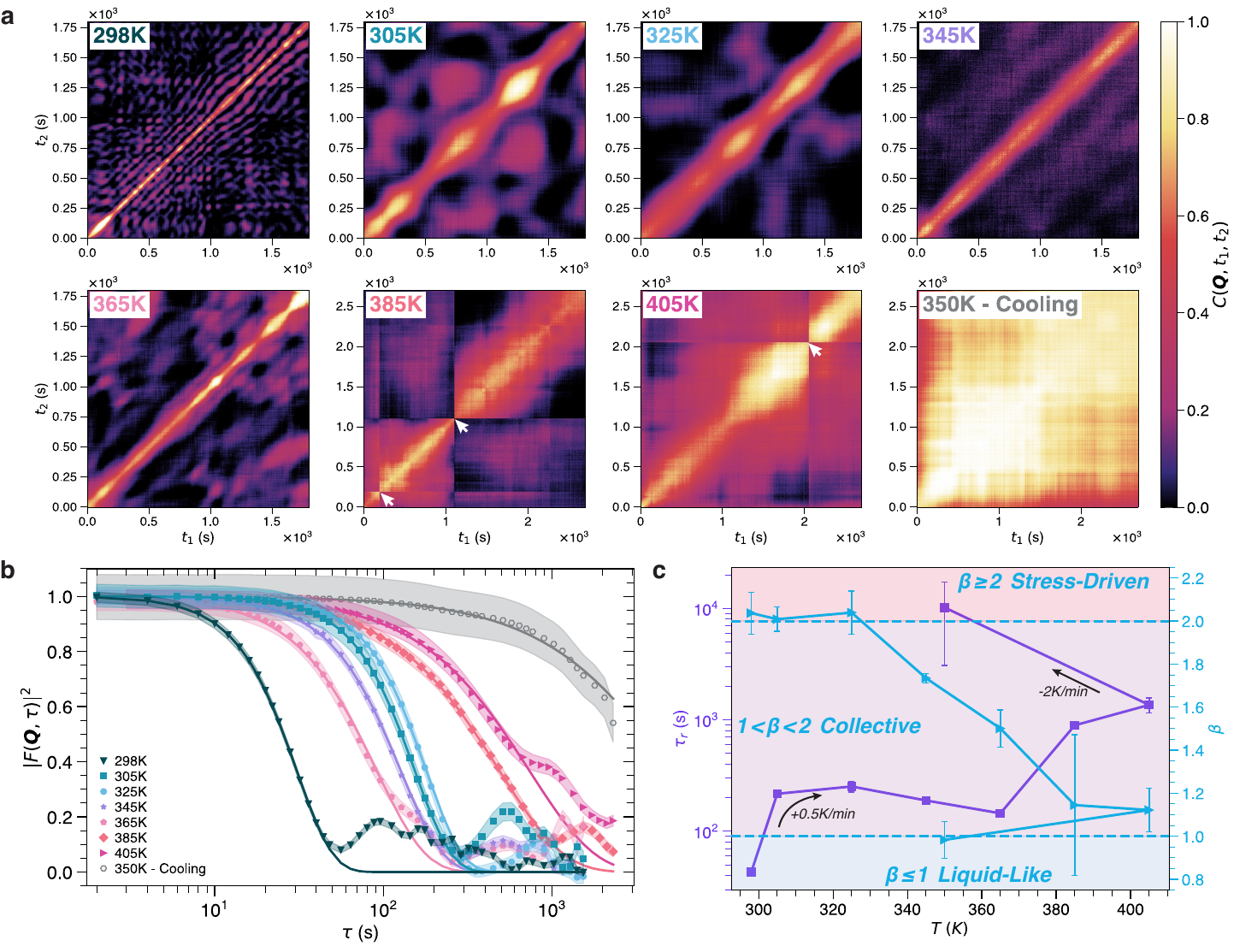}
    \caption{ \textbf{Annealing  Metastability in \bmNPS.}
    \textbf{a} \TTCFull matrices for the full temperature series \Tpath{298}{405}{350} (cooling pass). The \TTCFull matrices dramatically illustrate complex, dynamical heterogeneities that are not as evident in the inner product of the ISF \ISFIP, as well as rare, system-spanning avalanche events at times \tage = \qty{192}{\second} and \qty{1101}{\second} for \qty{385}{\kelvin} and \tage = \qty{2046}{\second} for \qty{405}{\kelvin} (white arrows). \textbf{b} ISFs \ISFIP calculated from the central mask partition via \gTwoFull multi-tau autocorrelation function, representing dynamics of the 211 peak for \bmNPS throughout the annealing protocol. Shaded bands show the standard deviation over pixels in the multi-tau calculation. At this reflection, dynamics are representative of local fluctuations at the boundary of tetragonal variant domains on the order of the correlation length \corrLength = \num{34.2} to \qty{54.7}{\nano\meter} in size, with resolution on the order of the lattice spacing for the reflection, \spacing = \spacingEqn = \num{2.856} to \qty{2.859}{\angstrom}. \textbf{c} Characteristic relaxation time \taur and shape parameter \shape extracted from fits to the intermediate scattering functions in \textbf{b} (solid lines), using (\ref{eq:g2}) and (\ref{eq:KWW}). Error bars are the standard deviation of fits to three independent time blocks of \TTCFull (Supplementary Materials).
}\label{fig:fig_g2_tau_beta}
\end{figure}

We tracked a single primary particle with XPCS across eight temperatures to resolve how thermal annealing alters the metastable strain-glass landscape in \bmNPS. Two-time correlations \TTCFull in Fig.~\ref{fig:fig_g2_tau_beta}\textbf{a} exhibit dynamical heterogeneities, including quasi-periodic and aging recorrelations discussed earlier, as well as system spanning avalanche events indicated by the white arrows. ISFs \ISFIP in Fig.~\ref{fig:fig_g2_tau_beta}\textbf{b} were derived from the observable autocorrelation functions \gTwoFull by fitting to equation~(\ref{eq:g2}) with the KWW exponential equation~(\ref{eq:KWW}), and are representative of the average dynamics at each temperature. The extracted relaxation times \taur in Fig.~\ref{fig:fig_g2_tau_beta}\textbf{c} (violet) obey a non-Arrhenius evolution suggesting a lack of well-defined energy scales, with dynamical processes slowing overall, but non-monotonically, as the temperature increases. 

Warming from \qty{298}{\kelvin} to \qty{305}{\kelvin} quintupled $\tau_r$ and expanded $\xi$ by \qty{35}{\percent} (Supplementary Materials, Fig. S2), demonstrating that even modest annealing coarsens the strain-glass domains, reducing the domain wall density available for rearrangement and slowing the dynamics. The relaxation time then gradually decreases from \qty{325}{\kelvin}, reaching a local minimum at \qty{365}{\kelvin} of $\taur\sim\qty{145}{\second}$, before increasing to $\sim$\qty{893}{\second} at \qty{385}{\kelvin}. We speculate that above \qty{365}{\kelvin} domain wall pinning barriers (such as disclination pairs or cubic phase impurities) are overcome, releasing the residual stresses accumulated during ball-milling in large, discrete events detected at \qty{385}{\kelvin} and \qty{405}{\kelvin}, after which the depleted stress reservoir no longer drives fast relaxation~\cite{sethna_deformation_2017}. A maximum temperature of \qty{405}{\kelvin} was reached in the series, well below reported structural phase transitions in \NPS beginning at roughly \qty{520}{\kelvin}~\cite{yu_na-ion_2016}, indicating the dynamics described correspond to the metastable cubic phase.  Upon cooling from \qty{405}{\kelvin} to \qty{350}{\kelvin}, the relaxation time increases by an order of magnitude from $\sim$$10^3$ to $\sim$$10^4$~\unit{\second} coinciding with an \qty{85}{\percent} increase in \corrLength relative to \qty{405}{\kelvin} and a shift in \Qnot from \qty{2.196}{\per\angstrom} to \qty{2.200}{\per\angstrom}, toward the tetragonal 211 position. A $\taur\sim\qty{e4}{\second}$ is beyond the measurement window and therefore indicates effectively arrested dynamics on the timescale of the measurement. Comparison of \taur at \qty{350}{\kelvin} with nearby temperatures \qty{345}{\kelvin} and \qty{365}{\kelvin} shows that domain wall relaxation on the cooling branch is roughly two orders of magnitude slower than on heating, indicating a strongly history dependent, non-reversible process. The slowdown between \qty{405}{\kelvin} and \qty{350}{\kelvin} corresponds to an apparent Arrhenius activation energy of $E_a = 0.45 \pm 0.16$~\unit{\electronvolt} (Methods), consistent with the thermally activated motion indicated by $\shape\sim1$ once the residual stress is depleted, whereas the difference from the heating branch at comparable temperature cannot arise from temperature alone and reflects the irreversible change in the defective microstructure. Effectively arrested dynamics at \qty{350}{\kelvin} also serve as a static reference for the sample environment throughout the temperature series \Tpath{298}{405}{350}, supporting the intrinsic origin of the measured hysteretic process.

We now turn to the trend of the shape parameter \shape in Fig.~\ref{fig:fig_g2_tau_beta}\textbf{c} (cyan), which characterizes the nature of the measured dynamics. An initial value of $\shape\sim\num{2}$ holds from \qtyrange{298}{325}{\kelvin}, indicating highly ``compressed" dynamics with stress-driven, constant velocity domain wall motion, as established in comparison with \htNPS~\cite{frey_liquid-like_2025}. Notably, this plateau persists through the initial five-fold increase in \taur, indicating that the driven character of the relaxation survives the early stages of coarsening and depinning. Above \qty{325}{\kelvin}, \shape falls monotonically, through collective dynamics at $\shape\sim\num{1.75}$ at \qty{345}{\kelvin} and $\shape\sim\num{1.5}$ at \qty{365}{\kelvin}, to $\shape\sim\num{1.1}$ at \qty{405}{\kelvin} approaching the simple exponential, thermally activated limit. In contrast to the strongly non-monotonic evolution of \taur, the monotonic decline of \shape indicates the two parameters report on distinct aspects of the relaxation, which we interpret as \shape tracking the progressive depletion of the residual stress driving the collective dynamics to individual liquid-like dynamics and \taur reflecting the evolving pinning landscape~\cite{riechers_intermittent_2024}. Upon cooling to \qty{350}{\kelvin}, $\shape\sim\num{1.0}$ reaches the liquid-like boundary, compared to $\shape\sim\num{1.75}$ at a similar temperature of \qty{345}{\kelvin} on the warming branch, again indicating a non-reversible, history dependent transformation. Consistently, the quasi-periodic recorrelations described earlier weaken and lose regularity above \qty{325}{\kelvin}, tracking the decline in \shape over the same range. Configurational memory and compressed dynamics ($\shape\geq\num{1.5}$) have also been observed together in Mott insulator devices~\cite{danna_jamming-controlled_2026}, matching the range $1.5\lesssim\shape\lesssim2$ over which we observe recorrelations, suggesting they share an underlying cause. Here we attribute both to metastable confinement stemming from residual milling stress and defect density, whose depletion on annealing removes them together.

\noindent\textbf{Scale-Free Avalanche Events}

\begin{figure}[ht!]
    \centering
    \includegraphics[scale=0.55]{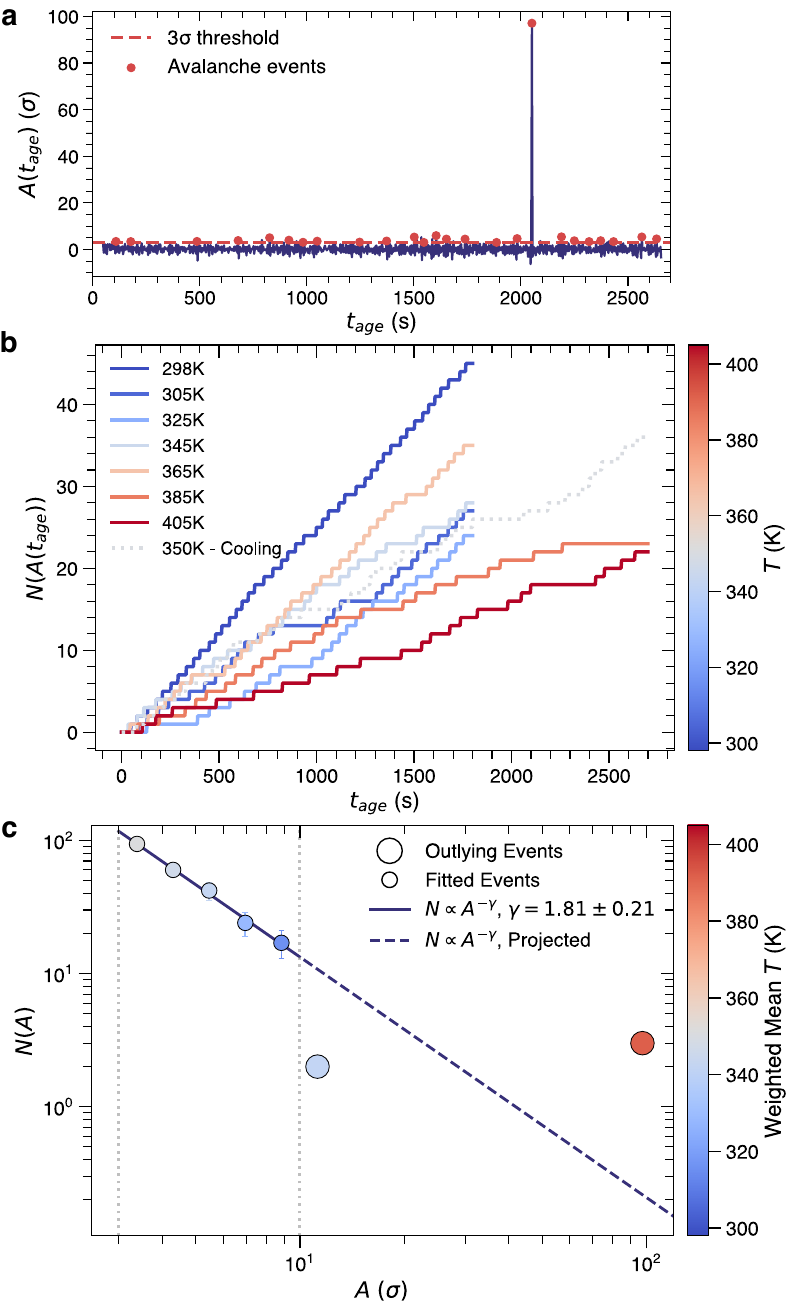}
    \caption{\textbf{Avalanches Follow Scale-Free Power Law.} \textbf{a} Representative avalanche detection $A(\tage)(\sigma)$, extracted from the two-time correlation matrix \TTCFull in Fig.~\ref{fig:fig_g2_tau_beta}\textbf{a} for temperature \qty{405}{\kelvin}, the maximum temperature reached in the cycle \Tpath{298}{405}{350} (cooling pass). The maximum avalanche event 97.06$\sigma$ above the noise floor is considered an outlying, system spanning event. \textbf{b} Cumulative avalanches $N(A(\tage))$ of all magnitudes irrespective of size detected with time, \tage, for  \TTCFull at each temperature. The roughly linear trend indicates no Omori law temporal clustering~\cite{sanborn_direct_2011} \textbf{c} Histogram of avalanche events $N(A)$ for all temperatures in the series. Marker colors represent the weighted mean temperature according to the respective number of avalanche events for each temperature contributing to the bin. The scale-free power law $N\propto A^{-\gamma}$ was fit to events between $3\sigma\leq A\leq9.96\sigma$, denoted by the dashed, vertical gray lines. }
    \label{fig:fig_avalanches}
\end{figure}

Intermittencies in the two-time correlation matrices \TTCFull, indicated by white arrows in Fig.~\ref{fig:fig_g2_tau_beta}\textbf{a}, are the signature of abrupt, cascading rearrangements of the nanoscale configuration occurring faster than the acquisition time of the XPCS measurement. Comparable intermittencies have been observed during aging in athermal, stress-driven, ferroelastic and strain glass systems~\cite{porta_intermittent_2018, ren_strain_2010,ren_strain_2014}. Interestingly, intermittencies are typically observed in ferroic systems under an applied field, whereas our measurements are isothermal with no other external stimuli, suggesting intermittencies in \bmNPS arise from stored internal stresses due to the mechanochemical synthesis, consistent with our $\shape>1$ interpretation. It has been suggested that intermittencies in ferroic matter are due to shear stress along the habit planes indicated in Fig.~\ref{fig:fig_intro}\textbf{b} and are potentially diffusive in origin~\cite{hurley_situ_2024,widera_non-ergodic_2015}. Of the three most dramatic intermittent events indicated, two occur at \qty{385}{\kelvin} and one at \qty{405}{\kelvin}, with the acquisition time for both temperatures being $\Delta t=\qty{3}{\second}$. However, closer inspection reveals all \TTCFull matrices show similar, albeit smaller, intermittent behavior along their diagonals $\tone= t_2$. This inspired a quantitative treatment of the intermittent, or ``avalanching," behavior observed by following the analysis techniques first demonstrated in~\cite{sanborn_direct_2011} and detailed in the Methods.

The extracted amplitude variations $A(\tage)$ of the correlation matrices are normalized to units of one-standard deviation $\sigma$ of the noise floor. Intuitively, $A(\tage)$ measures how rapidly the system's configuration is changing at time \tage, quantifying the change in the correlation profile between successive rows of \TTCFull, averaged over a window of $m$ slices on either side of the diagonal. A large $A(\tage)$ corresponds to a rapid loss of correlation, visible in the \TTCFull as a narrowing of the diagonal band, while a small $A(\tage)$ indicates the speckle pattern remains correlated within $m$. Avalanches are therefore identified as intervals in which the configuration rearranges much faster than the background fluctuation rate. As an example, we show $A(\tage)$ for \qty{405}{\kelvin} in Fig.~\ref{fig:fig_avalanches}\textbf{a}, where the large spike with a magnitude of $\sim$$97\sigma$ coincides with the white arrow position in the corresponding two-time correlation matrix. Smaller events occur at other \tage times with magnitudes near the $3\sigma$ threshold. Summing independent events with \tage for each temperature produces an avalanche activity plot in Fig.~\ref{fig:fig_avalanches}\textbf{b}, showing a linear relationship with \tage and the lack of an Omori law $\sim$$(1/\tage)^p$ scaling that would imply correlated, aftershock effects~\cite{sanborn_direct_2011}. The most active temperature is the initial, pristine measurement made at \qty{298}{\kelvin} for \bmNPS, followed by a decrease in activity from \qty{305}{\kelvin} to \qty{345}{\kelvin}. An increase in activity occurs at \qty{365}{\kelvin}, consistent with the non-monotonic trend observed in \taur in Fig.~\ref{fig:fig_g2_tau_beta}\textbf{c}, followed by another decrease from \qty{385}{\kelvin}--\qty{405}{\kelvin}.

Total avalanche statistics for all temperatures are shown in a logarithmically binned histogram in Fig.~\ref{fig:fig_avalanches}\textbf{c}, with the initial distribution $3\sigma<A<9.96\sigma$ fit to a scale-free power-law $N\propto A^{-\gamma}$, remarkably similar to the Gutenberg-Richter law for earthquake activity despite the vastly different scales involved~\cite{gutenberg_frequency_1944}. We found an exponent $\gamma = 1.81\pm0.21$, within the mutually shared error bounds for larger microstructural avalanches observed in ferroelastic martensites~\cite{sanborn_direct_2011,salje_ferroelastic_2012}, further supporting our picture of self-driven avalanches stemming from a quenched metastable strain-glass state in \bmNPS. Outlying events for $A>9.96\sigma$ do not scale, with two intermediate events from temperatures \num{298} and \qty{385}{\kelvin} lying below that predicted by the power-law and the three large events from temperatures \num{385} and \qty{405}{\kelvin} lying above what is predicted by the power-law. Intermediate events, with an avalanche magnitude of roughly 10$\sigma$, are similar in size to microstructural rearrangements observed in \ce{Co} using an equivalent analysis and measurement technique~\cite{sanborn_direct_2011}. Their scarcity here is therefore unlikely to reflect a detection limit or insufficient statistics, and instead suggests suppression by the finite size of the system at only $L\sim\qty{1.32}{\micro\meter}$. The three large avalanches with magnitude $\sim$$100\sigma$ occur only at high temperatures above \qty{365}{\kelvin}. They are considered \textit{system spanning} due to the total erasure of memory of the preceding nanoscale configuration implied by the block-like structures seen in \TTCFull (Fig~\ref{fig:fig_g2_tau_beta}\textbf{a}). That the three largest events lie above the extrapolated power law, and occur only above \qty{365}{\kelvin}, is consistent with the depletion picture developed above. Smaller, athermally driven rearrangements are exhausted at low temperature, leaving the most deeply pinned configurations, releasing only once sufficient thermal energy is available.

\section*{Discussion}\label{sec_discussion}

The observation of a cubic 211 Bragg peak with diffuse speckle from an isolated primary particle, $L\sim\qty{1.32}{\micro\meter}$ in size, provides direct single-particle confirmation that \bmNPS is crystalline, with globally cubic symmetry and local tetragonal motifs~\cite{krauskopf_local_2018,famprikis_under_2020}, rather than crystallites embedded in an amorphous matrix~\cite{hayashi_superionic_2012,hayashi_high_2014}. The dynamics of domain wall motion between tetragonal variants in the ferroelastic strain-glass framework proposed for \bmNPS can be understood through direct analogy to ferromagnetic or ferroelectric materials, with the deviatoric strain acting as the non-conserved order parameter in place of magnetization or polarization, bringing our understanding of SSEs within a broader family of well-established physical models under active investigation~\cite{ji_ferroic_2017,salje_ferroelastic_2012,shpyrko_direct_2007,mangu_hidden_2025}.
 
The enhanced ionic conductivity of cubic over tetragonal \NPS is widely attributed to local strain heterogeneities and elevated defect concentrations, given that both polymorphs exhibit similar local structures~\cite{krauskopf_local_2018,famprikis_fundamentals_2019,famprikis_under_2020,bo_structural_2016}. As the 211 reflection directly tracks local tetragonal distortions, our observed domain wall motion reveals a continuous reconfiguration of the local coordination environments that govern \ce{Na+}  migration bottlenecks and pathways~\cite{ahn_identification_2018,landgraf_disorder-mediated_2025}. Accelerated ionic transport is a hallmark of twin-wall boundaries in ferroelastic minerals~\cite{salje_fast_2000} and interfaces in \ce{Li+}-solid electrolytes~\cite{kwak_boosting_2023}. While cycling-induced defects create low-energy pathways that cause undesirable voltage fade in \ce{Li+}-rich cathodes, recoverable only via thermal annealing~\cite{singer_nucleation_2018,koh_advancing_2026}, our results show that annealing removes the defect-mediated dynamics in \bmNPS that may underlie its superior conductivity.
 
Although our measurement is not sensitive to \ce{Na+} directly, these observations suggest that the mesoscale domain structure and ionic transport may be coupled in \bmNPS and should not be treated as independent. Consistent with this, a \ce{Na+} random walk with the diffusivity implied by the room-temperature conductivity of \bmNPS would traverse the measured primary particle in an estimated \qtyrange{3}{10}{\second} (Methods), comparable to the relaxation times reported here. However, timescale agreement alone does not establish coupling. Determining whether stress-driven domain wall motion participates directly in ion migration will require pairing this measurement with a \ce{Na}-sensitive probe and transport data on identically annealed material. A systematic study of solid electrolytes with well-characterized broken local symmetries, correlated with their ionic conductivity, would establish the connection between cationic diffusion and lattice dynamics and advance defect engineering as a route to tuning transport in the SSE community.
 
In conclusion, Bragg XPCS resolves structural dynamics in \bmNPS on length and timescales inaccessible to established probes of ion transport, revealing a frustrated, stress-driven landscape that evolves on timescales relevant to long-range transport. The compressed relaxation, the memory-like recorrelations characteristic of non-ergodic glasses confined to metabasins, and the scale-free avalanches are distinct manifestations of a single strain-glass state. All three evolve together as residual milling stress is depleted on annealing, and all are absent in the tetragonal polymorph. Both the stress-driven dynamics \shape$\approx2$ and the enhanced conductivity of the ball-milled phase are removed by annealing, indicating that the mechanical stress state of the host lattice, embodied in a defect-rich microstructure of fine strain-variant domains, and not composition alone, is a variable available for tuning transport, identifiable by XPCS. This reframes mechanochemical synthesis as a strategy towards engineering defect and stress mediated landscapes rather than simply producing a metastable polymorph, and identifies frustrated, weakly pinned domain structures as a design target for solid-state electrolytes. 

\section*{Methods}\label{methods}
\noindent\textbf{Crystal Synthesis, Preparation, and Preliminary Characterization}
\bmNPS was synthesized via ball-milling a 75:25 molar mixture of \ce{Na2S} (Sigma Aldrich \qty{98}{\percent}) and \ce{P2S5} (Sigma Aldrich  \qty{99}{\percent}), respectively, for a minimum of two hours at 550 rpm. The sample was then flame sealed in a quartz capillary tube and heated to \SI{543}{\kelvin} at a rate of \SI{10}{\kelvin\per\minute} and held at temperature for two hours before being quenched in an ice bath with the intention of increasing the crystallite size while preserving the metastable cubic phase and its milling-induced defect population containing local tetragonal disorder \cite{nguyen_single-step_2019}. \htNPS was synthesized with the same initial ball-milling procedure as \bmNPS, then heated to \qty{772}{\kelvin} at \SI{10}{\kelvin\per\minute}, annealed for \qty{20}{\hour}, and allowed to cool to room temperature. The structural composition of \bmNPS and \htNPS was characterized via a Mo-source XRD instrument, the details and results of which are provided in Section S1 of the Supplementary Materials. 

For single primary particle Bragg XPCS characterization, dilute samples of \bmNPS and \htNPS were prepared by depositing raw SSE secondary particles in Loctite Stycast 2850FT. Stycast was prepared by first mixing with catalyst 9 in a weight ratio of 100:3.5, then immediately mixed with raw SSE powders in a Stycast-to-SSE weight ratio of approximately 100:0.3 before depositing in molds \qty{5}{\milli\meter} in diameter and \qty{0.75}{\milli\meter} in thickness, allowing cure at room temperature for \qty{24}{\hour}. Stycast-SSE composite samples were then sealed in \qty{0.09}{\milli\meter} thick Uline Dri-Shield Moisture Barrier bags for transport to PETRA III. All preparation and packaging of the samples was performed in an argon atmosphere glovebox. Due to the Bragg scattering geometry, Stycast does not contribute to the measured dynamics, with single SSE grains acting as a Fourier filter, confirmed by the comparatively static reference results in Fig.~\ref{fig:fig_tet}\textbf{b}. The Stycast was also not found to alter the structural composition or interfere with the scattering of the SSEs by comparison of the measured Bragg peaks at PETRA beamline P10 with the \ce{Mo} lab-source XRD measurements (Fig. S1, Supplementary Materials). \\

\noindent\textbf{XPCS Experimental Details}
X-ray photon correlation spectroscopy measurements were performed at the P10 Coherence Applications Beamline in hutch EH2 at the PETRA III synchrotron at Deutsches Elektronen-Synchrotron (DESY). The undulator-generated X-rays were monochromatized by a cryo-cooled channel-cut Si111 crystal monochromator at an energy of \qty{14}{\keV}. The coherent X-rays were focused to \qtyproduct{1.7 x 2.3}{\micro\meter} using \ce{Be} compound refractive lenses (CRLs). A $\num{2070}\times\num{2167}$ pixel DECTRIS EIGER X 4M detector was used with a pixel size of \qty{75}{\micro\meter} for a total field of view of $155\times\qty{165}{\milli\meter\squared}$. The maximum count rate was $1\times10^6$ counts/s/pixel. The sample to detector distance was fixed at \qty{5001}{\milli\meter}. A sample environment for transmission geometry provided by P10 was used as the temperature stage, with an accessible temperature range of \qtyrange[range-phrase = --]{253}{473}{\kelvin} controlled by a Lakeshore 340 controller, allowing for thermally driven measurements under vacuum. 

The temperature was increased at a ramp rate of \qty{0.5}{\kelvin\per\minute} and allowed to stabilize for approximately 20 minutes before beginning an XPCS scan. Temperatures \qtyrange{298}{365}{\kelvin} were measured for 900 frames with \qty{0.5}{\second} exposure and \qty{1.5}{\second} wait time ($\Delta t = \qty{2}{\second}$) for a total of \qty{1800}{\second}. Temperatures \qty{385}{\kelvin}, \qty{405}{\kelvin}, and the cooling pass temperature \qty{350}{\kelvin} were measured for 900 frames with a \qty{1.5}{\second} exposure and a \qty{1.5}{\second} wait time ($\Delta t = \qty{3}{\second}$) for a total measurement time of \qty{2700}{\second}. The measured relaxation times are one to four orders of magnitude longer than these acquisition rates, making these changes in $\Delta t$ inconsequential to \taur. We note that dynamics slow down at the temperatures where exposure is increased threefold. Beam induced dynamics from increased exposure time are expected to speed up relaxation times with increased dose, not to slow down, providing a check on stability and ruling out beam driven effects. Furthermore, we used an identical sample matrix, temperature stage, and coherent X-ray beam properties at the same beam-time for the cubic and tetragonal measurements. This confirms that the observed differences in symmetry dependent dynamics discussed in Results arise from the SSEs themselves, and are not beam induced or due to an experimental artifact. \\

\noindent\textbf{XPCS Analysis Details}
Coherent X-ray sensitivity to strain arises from atomic displacements $\mathbf{u}(\mathbf{r})$ through the infinitesimal strain $\epsilon_{ij} = \frac{1}{2}\left(\nabla\mathbf{u}(\mathbf{r}) + \nabla\mathbf{u}(\mathbf{r})^\mathsf{T} \right)$, which in a Bragg reflection appears as a phase shift $\phi = \Q \cdot \mathbf{u}(\mathbf{r})$~\cite{pfeifer_three-dimensional_2006,krivoglaz_x-ray_1996}. The cumulative phase shifts modulate the scattered intensity, $I \propto |\int_V s(\mathbf{r})e^{-i(\Q\cdot(\mathbf{r}+\mathbf{u}(\mathbf{r}))}dV|^2$, mathematically equivalent to the Fourier transform of the atomic form factor $s(\mathbf{r})$,  producing speckle~\cite{goodman_speckle_2020,krivoglaz_x-ray_1996,sutton_observation_1991}. Under Bragg condition, XPCS is sensitive to fluctuations within coherent volumes corresponding to the visible span of the speckle pattern, $\Delta \Q\approx 2\pi/\corrLength = \qty{2e-2}{\per\angstrom}$, with a time resolution equal to the detector frame spacing of $\Delta t$ \cite{shpyrko_direct_2007}.

The intensity-intensity autocorrelation function \gTwoFull is calculated as
\begin{equation}\label{eq:g2}
    \gTwoFull = \left\langle\frac{\langle I(\Q,t)I(\Q,t+\tau) \rangle_{t}}{\langle I(\Q,t) \rangle^{2}_{t}}\right\rangle_{p} = 1+A_0\ISFIP,
\end{equation}
where $I(\Q,t)$ and $I(\Q,t+\tau)$ are the intensities in a given pixel for frames taken at times $t$ and $t+\tau$. As shown in Fig.~\ref{fig:fig_xpcs}\textbf{b}, averaging is performed over all pairs of frames separated by $\tau$ in $t$, denoted by $\langle\cdot\rangle_{t}$, followed by a pixel average over $N_{p}$ pixels, denoted by $\langle\cdot\rangle_{p}$, within a region-of-interest to calculate a representative \gTwoFull for a given \Q. The error in \gTwoFull for each $\tau$ calculated as the standard deviation of \gTwoFull across all pixels divided by $\sqrt{N_p}$ (the standard error of the mean). $A_0$ is the contrast, a property that depends on beam coherence, geometry of the experiment, and properties of the sample. \ISFIP is the squared modulus of the normalized intermediate scattering function, related to \gTwoFull by the Siegert relation~\cite{siegert_fluctuations_1943}. An alternative method used to visualize XPCS dynamics is the mean-subtracted two-time correlation function \TTCFull, defined as
\begin{equation}
\begin{aligned}
\label{eq:C}
    \TTCFull &=\frac{\langle [I(\Q,t_1)-\langle I(\Q,t_1)\rangle_{p}][I(\Q,t_2)-\langle I(\Q,t_2)\rangle_{p}]\rangle_{p}}{\langle I(\Q,t_1)\rangle_{p}\langle I(\Q,t_2)\rangle_{p}} \\
    &= \frac{\langle I(\Q,t_1) I(\Q,t_2)\rangle_{p}}{\langle I(\Q,t_1)\rangle_{p}\langle I(\Q,t_2)\rangle_{p}} - 1.
\end{aligned}
\end{equation}
Fig.~\ref{fig:fig_xpcs}\textbf{c} shows \TTCFull for \bmNPS at \qty{298}{\kelvin} with the $(t_{age},\tau)$-coordinate system indicated at the center. A reparameterization of ($t_1$,$t_2$), $t_{age} = (t_1+t_2)/2$ is the time elapsed since the start of the measurement, and $\tau=|t_1-t_2|$ is the usual delay-time~\cite{bikondoa_use_2017}. \TTCFull can be projected onto the $\tau$-axis by averaging along the anti-diagonals, over $t_{age}$ at fixed $\tau$, as indicated in Fig.~\ref{fig:fig_xpcs}\textbf{c,d}. Thus 
\begin{equation}
    C(\Q,\tau) = \langle C(\Q,t_{age},\tau) \rangle_{t_{age}},
\end{equation}
and, by comparison with equations (\ref{eq:C}) and (\ref{eq:g2}), we find that $C(\Q,\tau)\propto\ISFIP$. As complementary visualizations, \gTwoFull and $C(\Q,\tau)$ contain equivalent dynamical information, yielding the same relaxation parameters, and differing only in overall contrast, which is reduced when the pixel average is performed before correlation~\cite{sutton_observation_1991,brugnara_time-resolved_2026}.

To normalize \gTwoFull calculations against changes in contrast stemming from the profile of the Bragg peak, the speckle pattern was partitioned into concentric elliptical rings, each containing an equal integrated intensity. For each scan, an elliptical profile was defined with semi-axes equal to the standard deviations $\sigma_x$ and $\sigma_y$ extracted from a two-dimensional Gaussian fit to the time-integrated intensity. The elliptical mask was then partitioned into five elliptical sub-rings by numerically integrating the intensity outward from the center until each ring contained \qty{20}{\percent} of the total intensity within the elliptical profile. The exception is for \htNPS, where a single partition with semi-axes equal to $2\sigma_x$ and $2\sigma_y$ standard deviations extracted from the Gaussian fits was used. This is due to the approximately five times narrower Bragg peak, functioning as a single speckle, making partitioning challenging without providing sufficient $\Delta \Q$ resolution.

The \gTwoFull was calculated pixel-by-pixel using a logarithmically $\tau$-binned calculation known as multi-tau, and averaged over all pixels within a given partition to represent the dynamics of that partition, as described in equation (\ref{eq:g2}). KWW fits using equation (\ref{eq:KWW}) were performed to the leading relaxation only, for each partition. Uncertainties for the fitted parameters \taur and \shape are derived from variation in KWW fits to independent block segments of \TTCFull with $t_{age}$. Values reported in the main text are from the central partition (violet, Fig.~\ref{fig:fig_xpcs}\textbf{b}) due to it possessing the highest signal-to-noise ratio and the \Q-dependence, fit parameters, and error estimation techniques are detailed further in Supplementary Materials. \\

\noindent\textbf{Avalanche Detection and Statistics}
Avalanche analysis was performed by following the methods detailed in~\cite{sanborn_direct_2011}. Amplitude variation $A(\Q,\tage)$ was extracted from the two-time correlation matrices \TTCFull by averaging over windows of size $m$ slices above and below the diagonal \tone= $t_2$. Quantitatively, this is
\begin{equation}
\begin{aligned}
    A_{\pm}(\Q,\tage) &= \frac{1}{m}\sum_{t_i=\tage\pm m}^{\tage\pm1}{[C(\Q,\tage,t_i)\mp C(\Q,\tage-1,t_i)]},\\
\end{aligned}
\end{equation}
with average
\begin{equation}
    A(\Q,\tage) =(A_{-}(\Q,\tage)+A_{+}(\Q,\tage))/2.
\end{equation}
For consistency with~\cite{sanborn_direct_2011}, a window size of $m=15$ was used. Due to the lack of a static reference sample, $A(\Q,\tage)$ was normalized to the standard deviation $\sigma$ of the lower 50th percentile of fluctuations in $A(\Q,\tage)$, considered the noise floor, for each temperature. Events were counted above a 3$\sigma$ cutoff and a minimum separation of \qty{15}{\second} in \tage, equal to the window size $m$, was used to distinguish distinct events. The histogram of avalanche magnitudes $N(A)$ used 15 logarithmically spaced bins in $A$, with error bars determined via Poisson statistics as $\sqrt{N(A)}$. \\

\noindent\textbf{Deviatoric Strain Tensor and Strain Glass Formalism}
To interpret the XPCS dynamics measured at the 211 Bragg peak for \bmNPS, it was necessary to identify the relevant order parameter and develop a formalism describing it. The following is largely inspired by the works of Khachaturyan~\cite{khachaturyan_2008}. For \bmNPS, the cubic-to-tetragonal distortion can be described by a stress-free eigenstrain tensor \eigenstrain
\begin{equation}
    \epsilon^0(1) = \begin{pmatrix}
    (c-a_0)/a_0 & 0 & 0\\
        0 & (a-a_0)/a_0 & 0\\
        0 & 0 &(a-a_0)/a_0
    \end{pmatrix},
\end{equation}
where $a_0=\qty{7.002}{\angstrom}$ is the cubic lattice parameter and $a=\qty{6.965}{\angstrom}$ and $c=\qty{7.114}{\angstrom}$ are the tetragonal lattice parameters, calculated from pseudo-Voigt fits in Fig.~\ref{fig:fig_intro}\textbf{c}, assuming a fully cubic-to-tetragonal distortion on the local level as an upper bound. For \eigenstrain, $p=(1,2,3)$ are cyclic permutations of the three tetragonal variants $c$-lattice parameter relative to the cubic basis. The eigenstrain can be decomposed into a dilatational and deviatoric tensor by \eigenstrain = \deviatoric+ \meanstrain\textit{I}, with the dilatational tensor \meanstrain\textit{I} being the mean of the diagonal elements in \eigenstrain, describing a homogeneous background strain. This gives $\meanstrain =\frac{1}{3}\Tr{\epsilon^0(p)} =((c+2a)/3a_0-1) = \num{1.9e-3} >0$, predicting a dilated volume for tetragonal distortions from cubic symmetry, indicating \bmNPS is denser than the purely tetragonal phase \htNPS. Densification may reflect elastic accommodation of neighboring variants \eigenstrain along the habit planes illustrated in Fig.~\ref{fig:fig_intro}\textbf{b}. The deviatoric strain \deviatoric can then be found as
\begin{equation}\label{eq:deviatoric}
    \epsilon'(1) = \epsilon^0(1)-\meanstrain = \frac{2(c-a)}{3a_0}\begin{pmatrix}
    1 & 0 & 0\\
        0 & -\frac{1}{2} & 0\\
        0 & 0 &-\frac{1}{2}
    \end{pmatrix}
\end{equation}
likewise for permutations $p=2,3$. Thus, the traceless deviatoric component carries the symmetry-breaking distortion and the elastic interactions between variants, while the isotropic dilatational residual changes only the average cell volume and would not produce significant contrast fluctuations in the measured coherent diffraction. Furthermore, the multiplicative factor characterizing the relative amplitude of deviatoric strain is $2(c-a)/3a_0 = \num{1.4e-2}$, an order of magnitude larger than the dilatational component \meanstrain\xspace and the dominant source of strain in the system. As \eigenstrain\xspace is defined relative to the fully distorted tetragonal cell, \num{1.4e-2} represents an upper bound on the local deviatoric amplitude in \bmNPS. A variant embedded coherently among the other two carries a constrained strain reduced from the stress-free value by elastic accommodation across the habit planes of Fig.~\ref{fig:fig_intro}\textbf{b}, with the residual concentrated near domain walls and triple junctions. The near-cubic metric of the average cell reflects the exact cancellation $\langle\deviatoric\rangle_V=0$ for equal variant populations and is therefore independent of this amplitude. We then choose to define the order parameter for local nanoscale fluctuations in \bmNPS as a linear combination of deviatoric strain variants, with the total deviatoric strain at point $\mathbf{r}$ as
\begin{equation}\label{eq:total_strain}
    \strain = \sum_{p=1,2,3} \eta_p(\mathbf{r})\deviatoric
\end{equation}
where $\eta_p(\mathbf{r})\in [0,1]$ and $\sum_{p=1,2,3}\eta_p(\mathbf{r})\leq1$. The strain variant fraction $\eta_p(\mathbf{r})$ is not conserved and the condition that the sum is less than or equal to one allows for a residual cubic parent phase or defect contributions. If the populations of the three variants are equal within a volume $V$, then $\langle \strain\rangle_V = \frac{1}{V}\int\strain dV = 0$, and the average deviatoric strain vanishes globally. In the pristine \bmNPS phase at \qty{298}{\kelvin}, the broad cubic 211 peak (Fig.~\ref{fig:fig_intro}\textbf{c}) suggests a local structure that is composed of deviatorically strained, tetragonal domains of roughly equal population such that $\langle \eta_p(\mathbf{r})\eta_p(\mathbf{r}+\mathbf{R})\rangle\sim e^{-|\mathbf{R}|/\corrLength}$~\cite{Chaikin_Lubensky_1995}. The resulting correlation length \corrLength = \num{34.2}$\pm$\num{0.5}~\unit{\nano\meter} is interpreted as the mean coherent domain size of each tetragonal variant, consistent with published values~\cite{famprikis_under_2020,krauskopf_local_2018}. \\

\noindent\textbf{Apparent Activation Energy on Cooling}
Assuming Arrhenius behavior on the cooling branch, $\tau_r = \tau_0 \exp(E_a/k_B T)$, the relaxation times at \qty{405}{\kelvin}  and \qty{350}{\kelvin} give
\begin{equation}
  E_a = \frac{k_B \ln(\tau_{350}/\tau_{405})}{1/350 - 1/405} = 0.45 \pm 0.16~\mathrm{eV},  
\end{equation}
using $\tau_{405} = 1361 \pm 213$~\unit{\second} and $\tau_{350} = (1.02 \pm 0.71)\times10^4$~\unit{\second} (Supplementary Materials, Table S3), with uncertainty propagated from both. Because $\tau_{350}$ exceeds the fit range and only two temperatures are available, this is an apparent activation energy under an Arrhenius assumption rather than a test of Arrhenius behavior. \\

\noindent\textbf{Ionic Transport Timescale Estimate} 
The relaxation times measured here, $\taur\sim\qty{43}{\second}$ for pristine \bmNPS at \qty{298}{\kelvin}, are many orders of magnitude slower than single-site \ce{Na+} hopping. The relevant comparison for long-range transport, however, is not the hop rate but the time required for an ion to traverse the illuminated particle. We estimate this from the Nernst-Einstein relation, $D_{\sigma}=\sigma k_B T/(nq^2)$~\cite{vargas-barbosa_dynamic_2020}. Taking the average cubic lattice parameter $a_0=\qty{7.002}{\angstrom}$ and assuming all six \ce{Na+} ions per unit cell are mobile gives a carrier density $n = 6/a_0^3 = \qty{1.75e22}{\per\cubic\centi\meter}$. With $q=\qty{1.602e-19}{\coulomb}$ and a room-temperature conductivity $\sigma\sim\qty{1e-4}{\siemens\per\centi\meter}$ for \bmNPS~\cite{famprikis_new_2019,krauskopf_local_2018,hogrefe_length-scale-dependent_2024}, this gives $D_{\sigma}=\qty{9.2e-10}{\centi\meter\squared\per\second}$. For three-dimensional random-walk diffusion the characteristic time to traverse a distance $L$ is $t=L^2/6D_{\sigma}$. The relevant length scale is the primary particle size, since the speckle pattern encodes the Fourier transform of the entire illuminated particle and, at \qty{298}{\kelvin}, $\shape\approx2$ indicates that \taur reports the collective relaxation of elastically coupled domains of size \corrLength within it~\cite{mangu_hidden_2025}. For the measured primary particle size $L\sim\qty{1.32}{\micro\meter}$ this gives $t\approx\qty{3.2}{\second}$.

This estimate carries uncertainty in both directions. A mobile fraction below unity would reduce $n$ and therefore increase $D_{\sigma}$, shortening $t$. Conversely, the Nernst-Einstein relation neglects ionic correlations, which reduce the true tracer diffusivity~\cite{marcolongo_ionic_2017}. The latter is captured by the Haven ratio $H_R$, with $D' = H_R D_{\sigma}$. Taking a representative Haven ratio indicative of highly collective diffusion of $H_R\sim0.3$ for illustration, the traversal time is increased to $t\approx\qty{10.5}{\second}$. Long-range ionic transport across the particle and collective rearrangement of the variant structure could therefore occur within roughly an order of magnitude of one another, rather than being separated by the many orders of magnitude that a single ionic hop would suggest. On the timescales relevant to long-range transport, the domain wall structure cannot be treated as static. \\

\backmatter

\noindent\textbf{Data Availability} The data supporting the findings of this study are available from the corresponding author upon reasonable request and will be deposited in a public repository upon publication.\\

\noindent\textbf{Code Availability} The analysis code used in this study is available from the corresponding author upon reasonable request.\\

\noindent\textbf{Supplementary Materials}
Supplementary Materials are available in the online version. \\

\noindent\textbf{Acknowledgements}
We thank A. Singer for illuminating discussions. This work is funded by the Energy Storage Research Alliance ``ESRA" (DE-AC02-06CH11357), an Energy Innovation Hub funded by the U.S. Department of Energy, Office of Science, Basic Energy Sciences. Efforts led by AF were supported by the National Science Foundation under Grant No. DMR-2145080. We acknowledge DESY (Hamburg, Germany), a member of the Helmholtz Association HGF, for the provision of experimental facilities. Parts of this research were carried out at PETRA beamline P10. Beamtime was allocated for proposals I-20221009 and I-20230664. \\

\noindent\textbf{Author Contributions}
Measuring nanoscale ionic diffusion mechanisms in solid-state electrolytes with XPCS was proposed by ESL, OGS, and YSM. Samples were synthesized and selected by PR. Synchrotron proposals were written by ESL, and all XPCS experiments were led by ESL. ND, NH, IZ, RG, and BKS assisted during the XPCS experiments. ZR, FW, and MS provided user support at PETRA beamline P10. ESL analyzed the data. ESL wrote the manuscript and interpreted the results with input from ND, JB, PR, EE, MW, YSM, and OGS. All authors contributed to the editing of the manuscript. AF, YSM, and OGS coordinated and obtained funding for this project.
\section*{Declarations}
The authors declare no competing interests.


\bibliography{sn_bibliography}

\end{document}


\title[Supplementary Material]{Supplementary Materials: Strain-Glass Dynamics in the Superionic Conductor \NPS}

\author*[1,2]{\fnm{Erik S.} \sur{Lamb}}\email{elamb@ucsd.edu}

\author[1,2]{\fnm{Nicol\`{o}} \sur{D'Anna}}

\author[3]{\fnm{Phillip} \sur{Ridley}}

\author[4,5]{\fnm{Nelson} \sur{Hua}}

\author[6]{\fnm{Ivan} \sur{Zaluzhnyy}}

\author[2,7]{\fnm{James} \sur{Burrow}}

\author[1,2]{\fnm{Emilio} \sur{Escauriza}}

\author[1]{\fnm{Robin} \sur{Glefke}}
\author[1]{\fnm{Boyan K.} \sur{Stoychev}}
\author[1,2]{\fnm{Mingze} \sur{Wan}}


\author[8]{\fnm{Zhe} \sur{Ren}}
\author[8]{\fnm{Fabian} \sur{Westermeier}}
\author[8]{\fnm{Michael} \sur{Sprung}}

\author[1]{\fnm{Alex} \sur{Fra\~{n}\'{o}}}

\author[2,3,7]{\fnm{Ying Shirley} \sur{Meng}}\email{shirleymeng@uchicago.edu}
\author*[1,2]{\fnm{Oleg G.} \sur{Shpyrko}}\email{oshpyrko@physics.ucsd.edu}

\affil*[1]{\orgdiv{Department of Physics}, \orgname{University of California San Diego}, \orgaddress{\street{9500 Gilman Dr}, \city{La Jolla}, \state{California}, \postcode{92093}, \country{United States of America}}}

\affil[2]{\orgdiv{Energy Storage Research Alliance}, \orgname{Argonne National Laboratory}, \orgaddress{\street{9700 South Cass Avenue}, \city{Lemont}, \state{Illinois}, \postcode{60439}, \country{United States of America}}}

\affil[3]{\orgdiv{Department of NanoEngineering}, \orgname{University of California San Diego}, \orgaddress{\street{9500 Gilman Dr}, \city{La Jolla}, \state{California}, \postcode{92093}, \country{United States of America}}}

\affil[4]{\orgdiv{PSI Center for Photon Science}, \orgname{Paul Scherrer Institute}, \orgaddress{\street{Forschungsstrasse 111}, \postcode{5232} \city{Villigen PSI}, \country{Switzerland}}}

\affil[5]{\orgdiv{Institute for Quantum Electronics}, \orgname{ETH Zürich}, \orgaddress{\postcode{8093} \city{Zürich}, \country{Switzerland}}}

\affil[6]{\orgdiv{Institute of Applied Physics}, \orgname{University of Tübingen}, \orgaddress{\street{Auf der Morgenstelle 10}, \postcode{72076} \city{Tübingen}, \country{Germany}}}

\affil[7]{\orgdiv{Pritzker School of Molecular Engineering}, \orgname{The University of Chicago}, \orgaddress{\street{5801 S Ellis Ave}, \city{Chicago}, \state{Illinois}, \postcode{60637}, \country{United States of America}}}

\affil[8]{\orgname{Deutsches Elektronen-Synchrotron DESY}, \orgaddress{\street{Notkestr. 85}, \postcode{22607} \city{Hamburg}, \country{Germany}}}

\maketitle

\tableofcontents
\clearpage


\section{Laboratory Powder Diffraction Characterization of the Pristine Samples}\label{snote:labxrd}

\begin{figure}[H]
	\centering
	\includegraphics[width=0.7\linewidth]{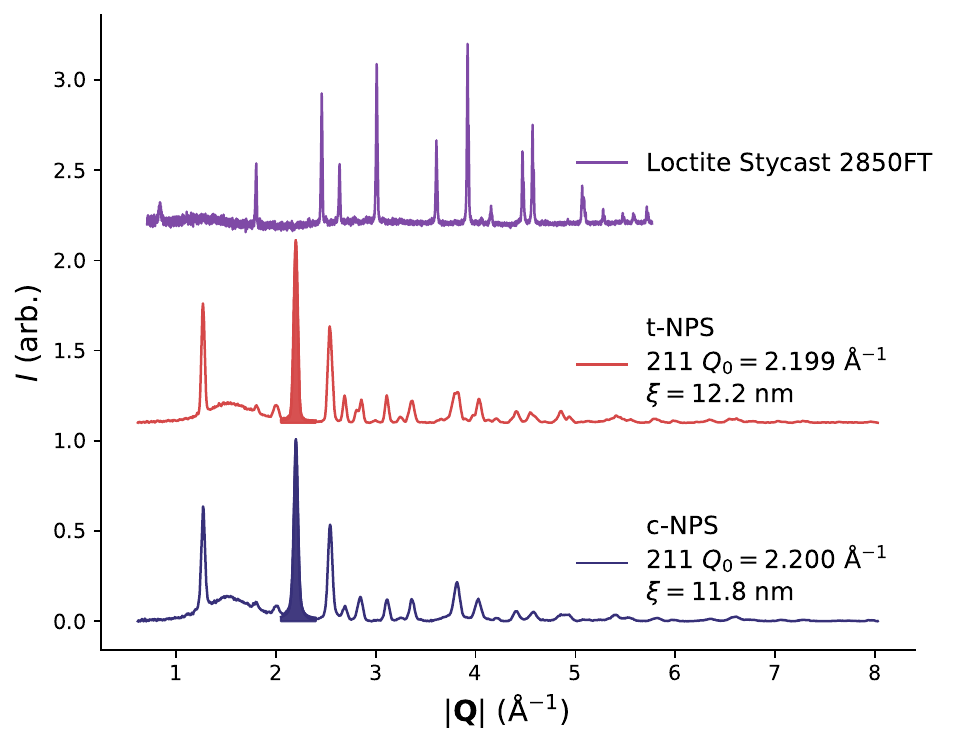}
	\caption{\textbf{Laboratory Powder Diffraction of \NPS Samples.} Normalized intensity as a function of $|\Q|$ for pristine \htNPS at \SI{298}{\kelvin} (red), pristine \bmNPS at \SI{298}{\kelvin} (blue), and Loctite Stycast 2850FT (purple) encapsulant used. Pseudo-Voigt fits to the \bmNPS and \htNPS 211 peaks are indicated by the filled profiles with fit results of \Qnot and \corrLength provided in the legends. Note that no Stycast peak overlaps with either of the SSE 211 peaks.}
	\label{sfig:labxrd}
\end{figure}
Prior to synchrotron measurements, both \bmNPS and \htNPS were characterized by laboratory powder diffraction to confirm phase identity and to provide an independent reference for peak positions and widths measured in the single-particle XPCS experiment. Solid electrolyte powders were loaded into 0.5 mm-diameter boron-rich glass capillary tubes (Charles Supper Company) while inside an Ar-filled glovebox. The tubes were sealed with clay, transferred out of the glovebox, and flame-sealed with a butane torch to create a hermetic seal. XRD patterns were collected over a 5–\qty{50}{\degree} 2$\theta$ range with an integration step size of \qty{0.01}{\degree} using a Bruker X8-ApexII CCD sealed tube diffractometer equipped with a molybdenum radiation source ($\lambda_{Mo} = \qty{0.7107}{\angstrom}$).

At the 211 reflection, the peak positions of pristine \bmNPS (\Qnot=\qty{2.200}{\per\angstrom}) and pristine \htNPS (\Qnot=\qty{2.199}{\per\angstrom}) at this reflection differ by less than \SI{0.05}{\percent} and do not by themselves distinguish the cubic and tetragonal phases. The Mo-source XRD instrument used introduces line broadening in transmission mode, causing several of the tetragonal peaks to overlap and appear as a single averaged peak. Consequently, this makes the correlation lengths \corrLength, shown in Fig.~\ref{sfig:labxrd}, extracted from the Bragg peak full-width at half-maximum of the Mo-source XRD profiles unreliable. Likewise, line broadening can cause overlap in the \bmNPS and \htNPS peak positions, making their peaks indistinguishable. Poor resolution stemming from instrumental line-broadening from the Mo-source XRD instrument used versus the higher resolution PETRA III synchrotron source easily accounts for discrepancies in \Qnot and \corrLength.

The 211 reflection was selected for the single-particle XPCS measurement for its intensity and sensitivity to individual tetragonal variants (Main, Fig.~1), not necessarily for its sensitivity to the cubic-to-tetragonal transition in a powder average, which are more apparent on the single, primary particle level. Their agreement with the single-particle 211 \Qnot measurements reported in the main text (Fig.~1 (Main) and Fig.~\ref{sfig:int_int_temp}) is within \qty{0.1}{\percent} and \qty{0.5}{\percent} for \bmNPS and \htNPS, respectively, and confirms that the Stycast-embedded single-grain samples used at the beamline are generally representative of the bulk powders from which they were drawn, despite the poor resolution of the lab source. Furthermore, a Rigaku \ce{Cu}
$K\alpha$--source XRD instrument was used to measure the line profile for cured Loctite Stycast 2850FT with catalyst 9, shown in Fig.~\ref{sfig:labxrd}, to ensure no peaks overlapped with the \NPS 211 peaks. The peaks in Loctite Stycast 2850FT appear consistent with $\alpha$--Alumina, included to improve thermal conductivity~\cite{delchamps1990stycast}. The nearest Stycast Bragg peak to the 211 \NPS peaks is located at \Q=\qty{2.460}{\per\angstrom} and is separated from the 211 peaks by approximately \qty{0.260}{\per\angstrom}, more than 10 times the visible span of the speckle pattern measured ($\Delta \Q\approx 2\pi/\corrLength =$\qty{2e-2}{\per\angstrom}, Main) at beamline P10, PETRA III. We therefore consider scattering from Stycast well separated and non-interfering with either of the \NPS peaks.

The 211 peaks were fit with pseudo-Voigt profiles, an interpolation between Lorentzian and Gaussian profiles defined as,
\begin{equation}\label{seq:psuedo_voigt}
V_p(|\Q|) = A[\eta L(|\Q|)+(1-\eta)G(|\Q|)] + c,
\end{equation}
where the Lorentzian $L(|\Q|)$ is defined as,
\begin{equation}\label{seq:lorentzian}
L(|\Q|) = \frac{1}{\pi}\frac{\gamma}{(|\Q|-\Qnot)^2+\gamma^2}, ~~\gamma = \frac{\text{FWHM}}{2},
\end{equation}
and the Gaussian is defined as, 
\begin{equation}\label{seq:gaussian}
G(|\Q|) = \frac{1}{\sigma\sqrt{2\pi}}\exp{\left(-\frac{(|\Q|-\Qnot)^2}{2\sigma^2} \right)}, ~~\sigma = \frac{\text{FWHM}}{2\sqrt{2\ln{2}}}.
\end{equation}
\Qnot is the peak position and FWHM is the Full-Width at Half-Maximum of the Bragg peak. The correlation length \corrLength we use throughout the Main and Supplemental is found through the Scherrer equation relevant for determining a lower-bound for coherently scattering domain size,
\begin{equation}\label{seq:scherrer}
    \corrLength = 2\pi K/\text{FWHM}.
\end{equation}
$K$ is the crystallite shape factor and is typically fixed close to unity, with the common choice of $K$=\num{0.9} used here. Equations (\ref{seq:psuedo_voigt})--(\ref{seq:scherrer}) were used in the subsequent Section~\ref{snote:bragg_peaks} to extract \Qnot and \corrLength from the integrated intensities measured from single primary particles at beamline P10, PETRA III.

\section{Temperature Dependence of Integrated Intensities from Synchrotron Source}\label{snote:bragg_peaks}

\begin{figure}[H]
	\centering
	\includegraphics[scale=0.5]{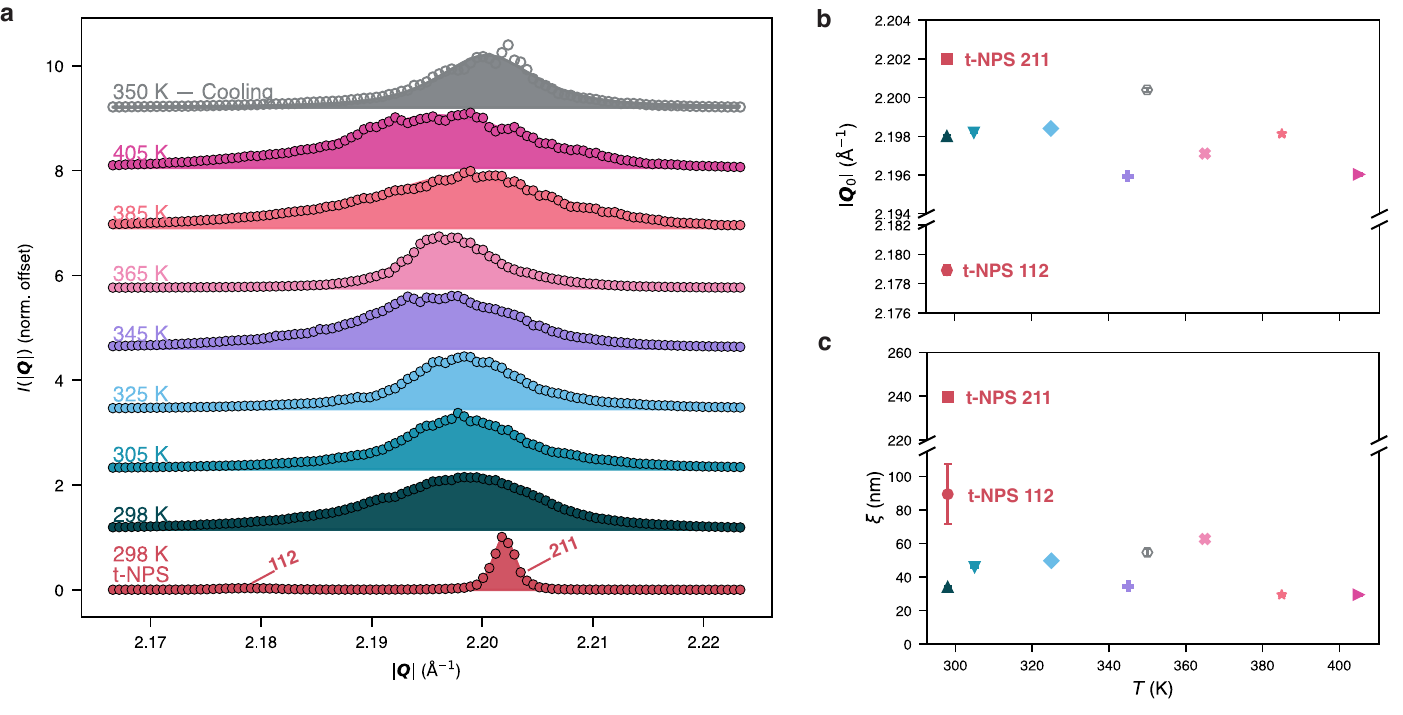}
	\caption{\textbf{Integrated Intensity Temperature Series.} \textbf{a} Azimuthally-\Q integrated intensities of the measured speckle patterns. Note that only the bottom most integrated intensity is for \htNPS, where the 112 and 211 reflections are indicated, and all others are for the 211 reflection for \bmNPS. The temperature series \Tpath{298}{405}{350} was made sequentially for \bmNPS with a \qty{0.5}{\kelvin\per\minute} warming rate and a \qty{2}{\kelvin\per\minute} cooling rate. Open markers are used for \qty{350}{\kelvin}-Cooling to indicate the measurement was made on a cooling pass. Pseudo-Voigt fits are indicated by the filled profiles.\textbf{b,c} Extracted \Qnot peak locations and correlation lengths \corrLength from the pseudo-Voigt fits. Note that \htNPS 112 and 211 peaks are indicated and \qty{350}{\kelvin}-Cooling has an open marker. All errors are statistical.}
	\label{sfig:int_int_temp}
\end{figure}

To confirm structural integrity and extract physically relevant information from \bmNPS and \htNPS Bragg reflections, such as the peak position \Qnot and the correlation length \corrLength, we projected the measured temporally averaged, two-dimensional Bragg spot onto the $|\Q|$-axis through azimuthal integration. Round markers in Fig.~\ref{sfig:int_int_temp}\textbf{a} indicate the integrated raw data, and the temporally averaged profiles offer a self-consistent check on measured dynamics by observation of ``jagged" profiles in later scans (at higher temperatures and during the subsequent cooling), indicating speckles that have not decayed and therefore producing slow dynamics. This is in contrast with the fast dynamics and correspondingly smooth profiles for early scans, such as at \qty{298}{\kelvin}.

There is no consistent trend in \Qnot or \corrLength with temperature for \bmNPS until the unambiguous shift in \Qnot towards the 211 peaks for \htNPS upon cooling to \qty{350}{\kelvin}. The potential for misalignment in the rocking curves made prior to each XPCS measurements cannot be neglected and the single particle nature of the experiment makes statistics inherently poor. Nonetheless, it is consistent with \bmNPS remaining in its cubic phase upon warming, and finally annealing towards a more enthalpically preferred, more tetragonal phase upon cooling to \qty{350}{\kelvin}, as described in Main. 

\section{Phase Sensitivity at the Cubic 211 Reflection}\label{snote:strain_phase}
In the weak scattering limit, the amplitude scattered by a domain displaced from its reference lattice position by a uniform field $\mathbf{u}$ acquires a phase factor $e^{i\mathbf{Q}\cdot\mathbf{u}} =e^{i\phi } $, where $\mathbf{Q}$ is the reciprocal lattice vector of the probed reflection \cite{krivoglaz_x-ray_1996,robinson_coherent_2009}. Importantly, $\phi =\mathbf{Q}\cdot\mathbf{u} $ is determined solely by the geometry of the strain and the scattering vector and is independent of the atomic form factors. \bmNPS adopts a cubic \emph{average} structure that is decomposed locally into three symmetry-equivalent tetragonal variants distinguished by the orientation of the elongated $c$-axis. In the formalism described in the Main, the three tetragonal variants are equivalent to the deviatoric strain order-parameter $\epsilon'(1)$ ($c \parallel [100]$), $\epsilon'(2)$ ($c \parallel [010]$), and $\epsilon'(3)$ ($c \parallel [001]$). From the positions of the 211 Bragg reflection for both \bmNPS and \htNPS and the 112 reflection for \htNPS, we extract
\begin{equation}\label{seq:lattice_params}
    a = 6.965\;\text{\AA},\quad c = 7.114\;\text{\AA},\quad a_0 = 7.002\;\text{\AA},
\end{equation}
giving $\Delta c \equiv c - a_0 = +\qty{0.112}{\angstrom}$ and $\Delta a \equiv a - a_0 = -\qty{0.037}{\angstrom}$ (tetragonal distortion $\eta = (c-a)/a_0 \approx \qty{2.1}{\percent}$). The uniform displacement of a unit cell belonging to variant $i$ relative to the cubic reference lattice is
\begin{align}
    \mathbf{u}_1 &= (\Delta c,\; \Delta a,\; \Delta a),\label{seq:u1}\\
    \mathbf{u}_2 &= (\Delta a,\; \Delta c,\; \Delta a),\label{seq:u2}\\
    \mathbf{u}_3 &= (\Delta a,\; \Delta a,\; \Delta c).\label{seq:u3}
\end{align}
The reciprocal lattice vector for the cubic 211 reflection is $\mathbf{Q}_{211} = (2\pi/a_0)(2,1,1)$. Forming $\phi_i = \mathbf{Q}_{211}\cdot\mathbf{u}_i$ for each variant:
\begin{align}
    \phi_1 &= \frac{2\pi}{a_0}\bigl(2\Delta c + \Delta a + \Delta a\bigr) = \frac{2\pi}{a_0}(2\Delta c + 2\Delta a),\label{seq:phi1}\\[4pt]
    \phi_2 &= \frac{2\pi}{a_0}\bigl(2\Delta a + \Delta c + \Delta a\bigr) = \frac{2\pi}{a_0}(3\Delta a + \Delta c),\label{seq:phi2}\\[4pt]
    \phi_3 &= \frac{2\pi}{a_0}\bigl(2\Delta a + \Delta a + \Delta c\bigr) = \frac{2\pi}{a_0}(3\Delta a + \Delta c).\label{seq:phi3}
\end{align}
The degeneracy $\phi_2 = \phi_3$ is \emph{exact}, following directly from $k = l$ in the Miller indices $(h,k,l) = (2,1,1)$, independently of the numerical values of $\Delta c$ and $\Delta a$. Substituting the measured lattice parameters gives,
\begin{equation}\label{seq:phases}
    \phi_1 \approx \qty{0.134}{\radian}, \qquad \phi_2 = \phi_3 \approx \qty{0.001}{\radian}.
\end{equation}
These per unit-cell phase shifts are summarized in Table~\ref{tab:phase} below.

\begin{table}[ht]
\centering
\caption{Phase shifts $\phi_i = \mathbf{Q}_{211}\cdot\mathbf{u}_i$ for each tetragonal variant evaluated at the cubic 211 reciprocal lattice vector, with $a_0 = 7.002$\,\AA, $\Delta c = +0.112$\,\AA, $\Delta a = -0.037$\,\AA.}
\label{tab:phase}
\begin{tabular}{lccc}
\toprule
Variant & Unique axis & $\phi_i$ (symbolic) & $\phi_i$ \\
\midrule
$\epsilon'(1)$ & $c \parallel [100]$ & $\dfrac{2\pi}{a_0}(2\Delta c + 2\Delta a)$ & $0.04265\pi$ \\[8pt]
$\epsilon'(2)$ & $c \parallel [010]$ & $\dfrac{2\pi}{a_0}(3\Delta a + \Delta c)$ & $0.000$ \\[8pt]
$\epsilon'(3)$ & $c \parallel [001]$ & $\dfrac{2\pi}{a_0}(3\Delta a + \Delta c)$ & $0.000$ \\
\bottomrule
\end{tabular}
\end{table}

\section{Primary Particle Size Extracted from Speckle Size}\label{snote:particle_size}
The characteristic primary particle size of the ball-milled \bmNPS and \htNPS samples were estimated from the spatial speckle size on the detector, using the Rayleigh criterion to relate the diffraction-limited speckle width to the size of the coherent scatterer. Because the transverse coherence length of the X-ray beam at PETRA P10 exceeds the ball-milled particle size, the speckle pattern is set by diffraction from individual particles acting as apertures.

\begin{figure}[ht]
	\centering
	\includegraphics[width=\linewidth]{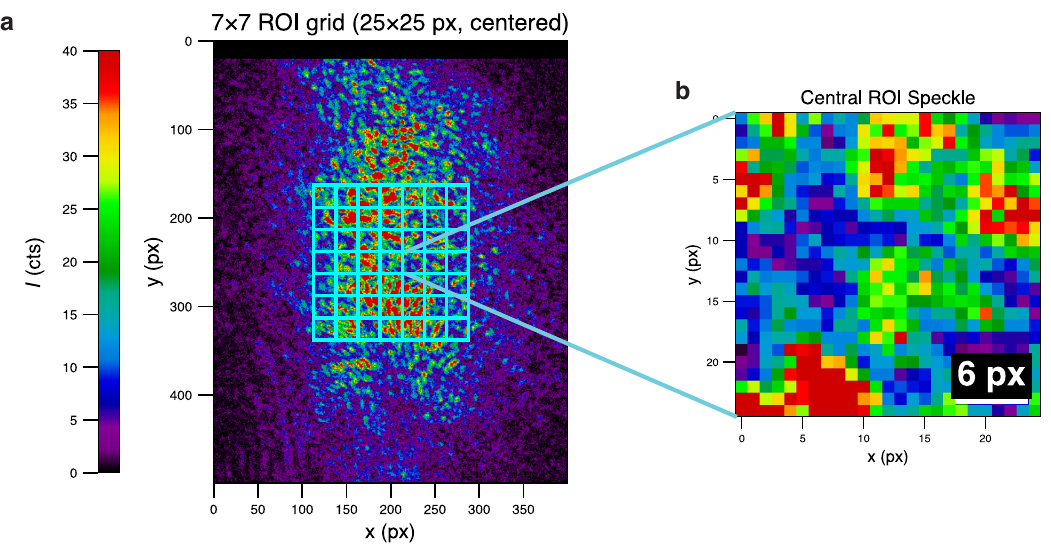}
	\caption{\textbf{\bmNPS Speckle Partitioning.} \textbf{a} Summed speckle pattern of the first 10 frames of the $211$ Bragg peak of \bmNPS at \qty{298}{\kelvin}. Overlayed is a $7\times 7$ grid (cyan) of $25\times 25$ pixel regions of interest (ROI), centered on the Bragg peak. \textbf{b} The central ROI is shown in the call-out to the right, with a scale bar indicating the characteristic speckle size in pixels of $\sim6$ pixels. For the DECTRIS EIGER X 4M detector used at PETRA P10, pixels are \qtyproduct{75x75}{\micro\meter} in size.}
	\label{sfig:speckle_partition}
\end{figure}

To obtain a statistical estimate of the speckle size across the Bragg peak, we summed the first ten frames of the $211$ Bragg reflection at \qty{298}{\kelvin} (Fig.~\ref{sfig:speckle_partition}\textbf{a}) and divided the central region into a $7\times 7$ grid of $25\times 25$ pixel independent regions of interest (ROIs), centered on the peak. Each ROI is small enough to sample a locally uniform region of the Bragg peak while containing many speckles for statistical averaging. A representative central ROI is shown in Fig.~\ref{sfig:speckle_partition}\textbf{b}. The detector at beamline P10 is a DECTRIS EIGER~X 4M with pixel size p = \qtyproduct{75x75}{\micro\meter}.

\begin{figure}[ht]
	\centering
	\includegraphics[width=\linewidth]{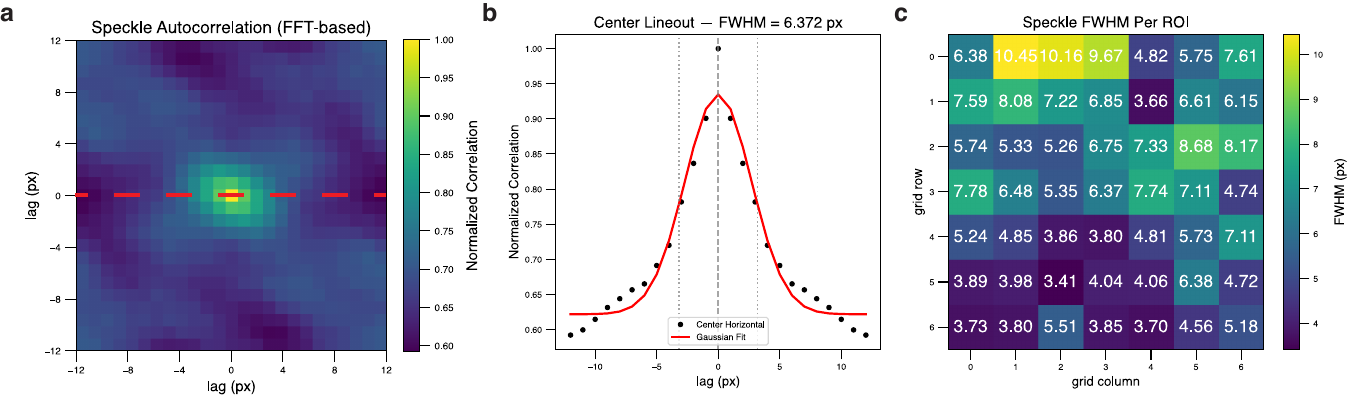}
	\caption{\textbf{Speckle-Size Autocorrelation Pipeline.} \textbf{a} Speckle spatial autocorrelation $X(\Delta x, \Delta y)$ for the central ROI shown in Fig.~\ref{sfig:speckle_partition}\textbf{b} calculated using equation (\ref{seq:cross_corr}). \textbf{b} Line-cut of the speckle spatial autocorrelation in \textbf{a} for $\Delta y = 0$ indicated by the red dashed line in \textbf{a}. A Gaussian fit using equation (\ref{seq:spatial_gauss}) is indicated by the red profile. The full-width at half-maximum (FWHM) from the fit is indicated in the title as being \num{6.372} px. \textbf{c} FWHM results for every ROI in Fig.~\ref{sfig:speckle_partition}\textbf{a} using the procedure detailed in \textbf{a,b}.  }
	\label{sfig:speckle_corr}
\end{figure}

For each ROI, the spatial autocorrelation of the speckle pattern was computed via the Fast Fourier Transform (FFT) as 
\begin{equation}\label{seq:cross_corr}
    X(\Delta x, \Delta y) = \mathcal{F}^{-1}\left[\,|\mathcal{F}\{I(x,y)\}|^{2}\,\right],
\end{equation}
where $I(x,y)$ is the ROI intensity, $\mathcal{F}$ denotes the 2D FFT, and the result is shifted to place zero lag at the array center and normalized so that $X(0,0)=1$. The autocorrelation peak width reflects the characteristic speckle size; the baseline reflects the average cross-correlation between distinct speckles. A representative autocorrelation for the central ROI is shown in Fig.~\ref{sfig:speckle_corr}\textbf{a}.

The central horizontal line-cut of $X(\Delta x, 0)$ (Fig.~\ref{sfig:speckle_corr}\textbf{b}) was fit to a Gaussian with a constant offset,
\begin{equation}\label{seq:spatial_gauss}
    X(\Delta x, 0) = A\,\exp\!\left[-\frac{(\Delta x - x_0)^2}{2\sigma^2}\right] + B,
\end{equation}
from which the full-width half-maximum was extracted as $\mathrm{FWHM} = 2\sqrt{2\ln 2}\,\sigma$. The fit parameters $A$, $x_0$, $\sigma$, and $B$ were constrained to physical ranges to ensure a positive width. The procedure was applied independently to each of the 49 ROIs, producing the spatial map of speckle FWHM shown in Fig.~\ref{sfig:speckle_corr}\textbf{c}. The mean speckle size is $\Delta\ell_{\mathrm{px}} = 5.92 \pm 0.67$ pixels, where the uncertainty is the standard deviation across the grid, divided by $\sqrt{N_{ROIs}}$, and reflects the irregular shape and internal heterogeneity of ball-milled particles.

Treating each particle as a diffraction-limited, circular aperture, the Rayleigh criterion relates the speckle size on the detector to the aperture (particle) size $L$ by
\begin{equation}\label{seq:rayleigh}
    L = 1.22\frac{D\,\lambda}{\Delta\ell_{\mathrm{px}}\, p},
\end{equation}
where $D = $~\qty{5.00}{\meter} is the sample-to-detector distance, $\lambda = \qty{0.89}{\angstrom}$ is the X-ray wavelength (corresponding to a photon energy of \qty{14}{\keV}), and $p = \qty{75}{\micro\meter}$ is the detector pixel size. Applying this conversion ROI-by-ROI, and averaging gives a characteristic particle size for \bmNPS of  $L$ = 1.33 $\pm$ \qty{0.06}{\micro\meter} at \qty{298}{\kelvin}. We use an ROI-by-ROI calculation of $L$ using equation (\ref{seq:rayleigh}) because of the physically meaningful heterogeneity in the speckle size observed in Fig.~\ref{sfig:speckle_corr}\textbf{c}. We repeat the same procedure for all other temperatures measured for \bmNPS, changing only the grid dimensions to accommodate narrower peaks and thereby filtering out background containing no speckle, only Poisson noise. 

\begin{table}[ht]
\centering
\caption{Compiled results for the average pixel size $\Delta\ell_{\mathrm{px}}$ and characteristic primary particle size $L$, the mean of the ROI-by-ROI calculation using equation (\ref{seq:rayleigh}), found for each temperature for \bmNPS using the methods described above. Also included are the ROI grid layouts used, where each ROI remained 25$\times$25 pixels in size. Smaller grids were used at later temperatures to ensure ROIs contained speckle and not background. }
\label{tab:speckle_sizes}
\begin{tabular}{lccc}
\toprule
$T$ $(K)$ & Grid & $\Delta\ell_{\mathrm{px}}$ & $L$ (\unit{\micro\meter}) \\
\midrule
298 & $7\times7$ & $5.92\pm0.26$ & $1.33\pm0.06$  \\[8pt]
305 & $7\times7$ & $6.41\pm0.30$ & $1.28\pm0.08$ \\[8pt]
325 & $7\times7$ & $6.34\pm0.25$ & $1.23\pm0.05$  \\[8pt]
345 & $5\times5$ & $5.23\pm0.41$ & $1.66\pm0.17$ \\[8pt]
365 & $5\times5$ & $5.26\pm0.30$ & $1.48\pm0.08$  \\[8pt]
385 & $5\times5$ & $6.17\pm0.43$ & $1.45\pm0.20$ \\[8pt]
405 & $5\times5$ & $6.56\pm0.54$ & $1.56\pm0.28$   \\[8pt]
350-Cooling & $5\times5$ & $6.31\pm0.51$ & $1.32\pm0.09$  \\
\bottomrule
\end{tabular}
\end{table}
The measured particle sizes are all statistically equivalent, indicating no primary particle size changes with temperature that would be indicative of grain growth or ablation. A consistent primary particle size also serves as evidence that we are indeed tracking the same primary particle, and that all dynamics measured are intrinsic to nanoscale dynamics discussed in Main, not attributable to changes in the primary particle size or shape. Taking the weighted mean of $L$ across all temperatures in Table~\ref{tab:speckle_sizes} gives us a characteristic primary particle size for \bmNPS of
\begin{equation}\label{seq:particle_size}
    L_{cub} = 1.32\pm\qty{0.04}{\micro\meter}.
\end{equation}
This value sets an upper bound on a single coherent reorganization event and provides the length scale that limits the size of avalanches arising from intra-particle dynamics. Furthermore, given the coherent beam spot size of $\num{1.7}\times$~\qty{2.3}{\micro\meter} used at beamline P10, a particle size of \qty{1.32}{\micro\meter} ensures that all dynamics measured are intrinsic to a single coherent primary particle and are not due to inter-grain rearrangements. By an identical analysis procedure, we found the characteristic particle size for \htNPS to be $L_{tet} = 3.52\pm$\qty{0.38}{\micro\meter}, suggesting the \htNPS grain is about \qty{50}{\percent} larger than the transverse coherence length. Since this would mean the measured speckle pattern is no longer diffraction limited, we believe the value obtained for $L_{tet}$ is not accurate, but does suggest the measured particle is larger than the coherent volume.

\section{Q-Dependence in \gTwoFull across Temperatures}\label{snote:q_dependence}

\begin{figure}[H]
	\centering
	\includegraphics[width=\linewidth]{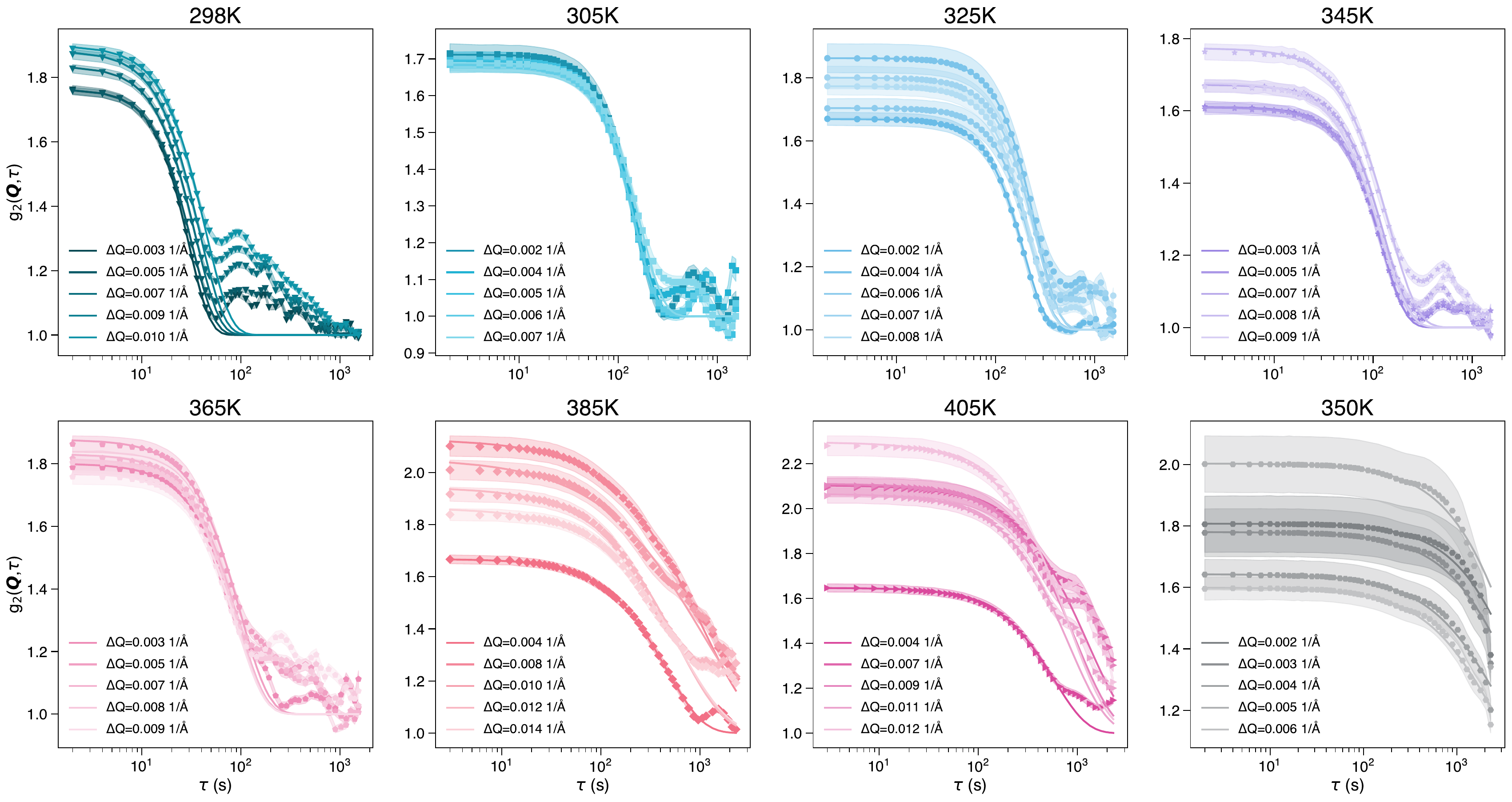}
	\caption{\textbf{\gTwoFull \Q-Dependence with Temperature.} Autocorrelation \gTwoFull for every equal integrated intensity partition, calculated according to Methods, with the minimum $\Delta\Q$ being the representative \gTwoFull for a given scan due to it containing the highest intensity and therefore the best signal-to-noise \cite{falus_optimizing_2006}. }
	\label{sfig:g2_dq}
\end{figure}

We chose only the central partition containing the minimum $\Delta\Q$ as representative of the dynamics for a particular temperature. As the central partition contains the most intense speckles, it has the best signal-to-noise ratio which goes as $\propto I^2$ for XPCS \cite{falus_optimizing_2006}. Due to the single particle nature of our scattering geometry, other scattering statistics arising from the powder ring and fringes or truncation rods scattered from the shape or facets of the particle could also influence measured dynamics for high $\Delta\Q$ partitions \cite{goodman_speckle_2020}. The central partition, therefore, is more insulated from these features and better represents the intrinsic nanoscale dynamics of the system. Non-speckle scattering influencing the measured dynamics at high $\Delta\Q$ also serves to explain the lack of any expected $\Delta\Q^{-1}$ (ballistic) or $\Delta\Q^{-2}$ (diffusive) scaling dependence and the generally non-monotonic behavior as a function of $\Delta\Q$ seen in Fig.~\ref{sfig:g2_dq}.

\begin{figure}[H]
	\centering
	\includegraphics[width=\linewidth]{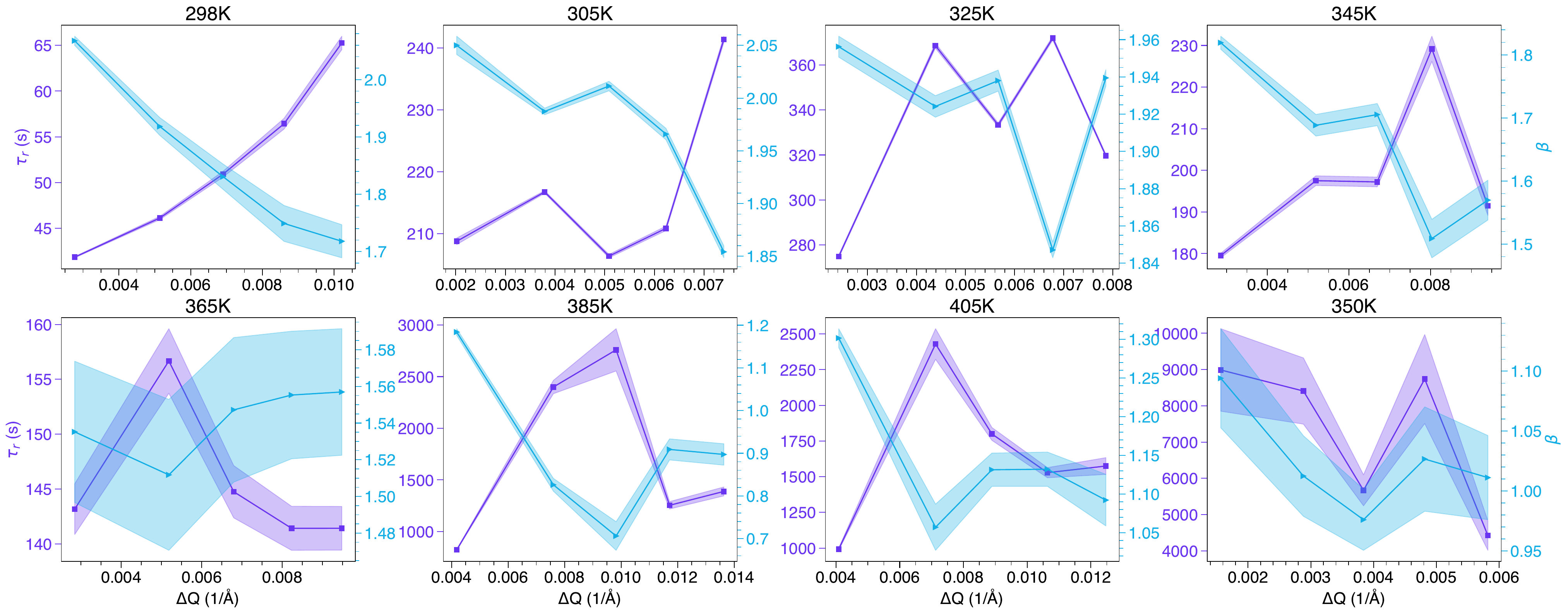}
	\caption{\textbf{\Q-dependent Dynamical Heterogeneity.} Relaxation time \taur (violet) and characteristic shape parameter \shape (cyan) as a function of $\Delta\Q$ for every temperature. While dynamical heterogeneity is present, there is no reliable $\Delta\Q^{-1}$ (ballistic) or $\Delta\Q^{-2}$ (diffusive) dependence. Variation versus $\Delta\Q$ could be due to additional contributions at high-$\Delta\Q$ stemming from fringes, particle shape, and a powder ring. }
	\label{sfig:taubeta_dq}
\end{figure}

\section{Block Error Estimation and Fit Parameters}\label{snote:partitions}

\begin{figure}[H]
	\centering
	\includegraphics[width=\linewidth]{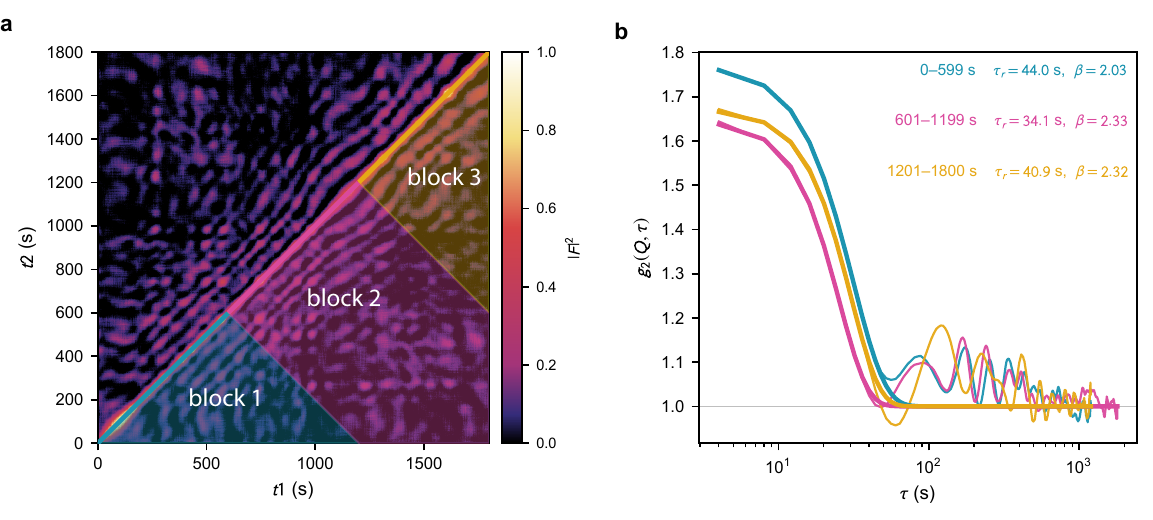}
	\caption{\textbf{Independent Block Error Estimate.} \textbf{a} Uncertainties for \taur and \shape were estimated by subdividing the two-time correlation matrix \TTCFull into three independent blocks of equal length along $t_{age}$ \textbf{b} $C(\Q,\tau)$ calculated from line-out averages for each block with the fit results displayed in the legend. The single-exponential KWW form was used for consistency with the Main text.}
	\label{sfig:rt_block_est}
\end{figure}

Error estimation for \taur and \shape extracted by KWW fits to the full \gTwoFull calculated from the entire measurement time can be challenging, with careful error propagation usually intractable. Statistical errors from the covariance matrix from \texttt{scipy.optimize.curve\_fit} \cite{virtanen_scipy_2020} generally underestimate a process that is not Gaussian distributed, such as a compressed exponential KWW. In the case where there is no $\Delta \Q$ dependence, uncertainties in \taur and \shape can be estimated from the standard deviation of the fit results across partitions. Although we showed a lack of $\Delta \Q$ scaling for our measurements in Section~\ref{snote:q_dependence}, the heterogeneities across $\Delta \Q$ due to confounding statistics make even this reasonable error estimation method inapplicable.
\begin{figure}[ht]
	\centering	
    \includegraphics[width=\linewidth]{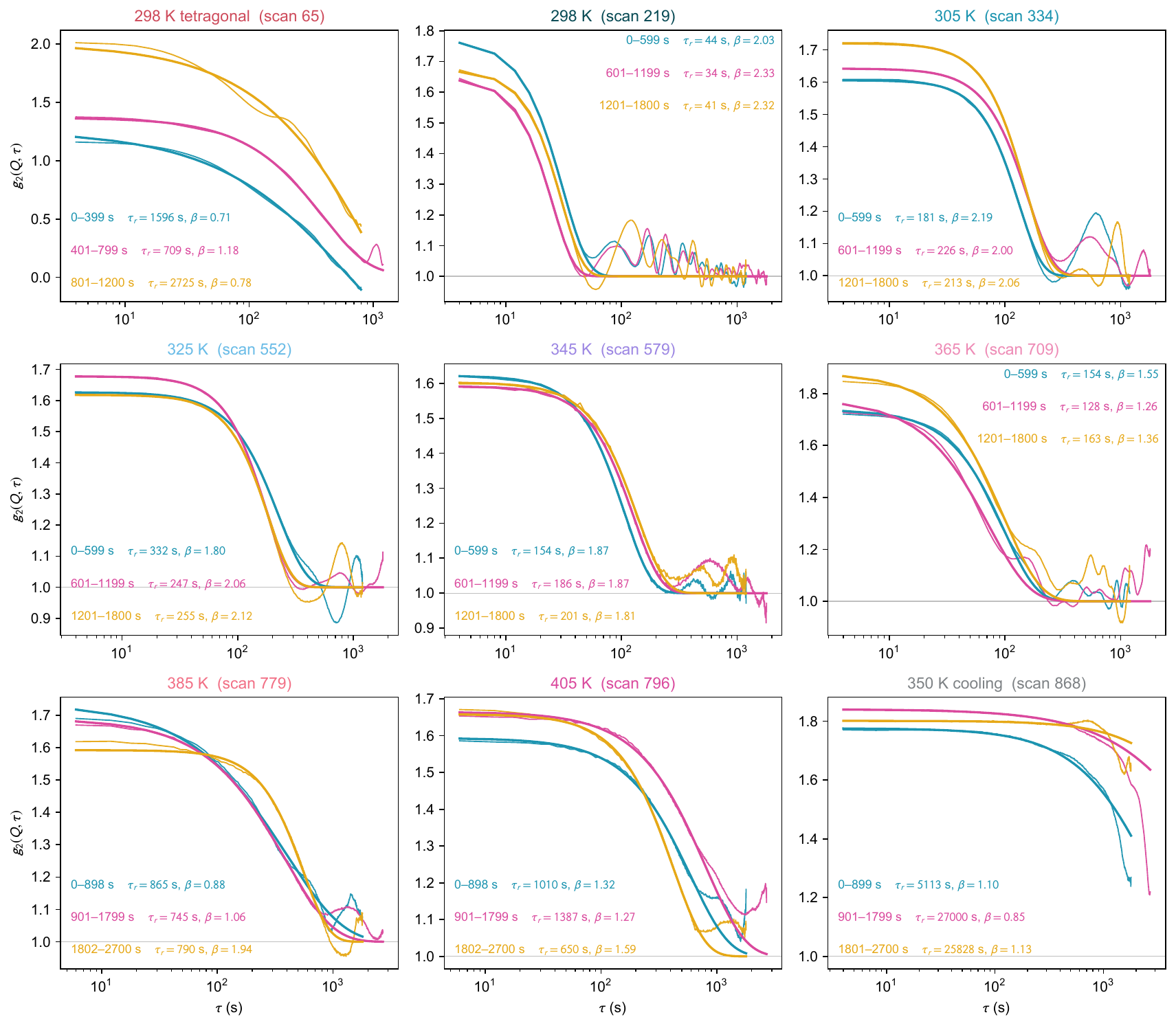}
	\caption{\textbf{Independent Block Fits per Temperature.} Block fit results using the same method as shown in Fig~\ref{sfig:rt_block_est}. All fits were performed with a KWW exponential and the Siegert relation with the exception of the measurement at \qty{298}{\kelvin} for \htNPS where we used a free baseline $c$ instead of \num{1}. This is due to the unrealistically high contrast stemming from the narrow, highly ordered Bragg peak measured for \htNPS.}
	\label{sfig:multi_block_est}
\end{figure}
For these reasons, we developed an alternative error estimation method we believe accurately represents the uncertainties in \taur and \shape parameters extracted from the KWW exponential. While the fit results from the full multi-tau \gTwoFull calculation from the central partition $\Delta \Q \approx 0$ for each temperature remain the representative values, the error is calculated by first dividing the two-time correlation matrices \TTCFull into three independent segments of equal length along \tage and calculating $C(\Q,\tau)$ for each segment, as shown in Fig.~\ref{sfig:rt_block_est}\textbf{a}. Each $C(\Q,\tau)$ is fit with the same KWW form as for the multi-tau fits (Fig.~\ref{sfig:rt_block_est}\textbf{b}) and the error for a given scan is the standard deviation in \taur and \shape across all three blocks divided by $\sqrt{3}$. We chose three independent blocks after finding that more than three led to $C(\Q,\tau)$ that were not properly normalized and fewer than three reduced our statistics. As shown in Fig.~\ref{sfig:multi_block_est}, this method was repeated across all temperatures for \bmNPS and for \htNPS at \qty{298}{\kelvin}, where a modified relation \gTwoFull = $c$ + $A$\ISFIP was used due to the unreasonably high contrast stemming from the sharp, highly-ordered 211 peak. We believe this method of estimating errors for dynamical parameters is appropriate and accurately reflects the uncertainty stemming from the measurement window and how dynamics evolve with \tage, thereby containing valuable additional information beyond a simple uncertainty.

The numerical fit results for the central $\Delta \Q \approx 0$ partition for \bmNPS and their errors estimated using the method above are displayed in Table~\ref{tab:bm_nps_fits} below. Fits were performed using the  Intermediate Scattering Function (ISF) $|F(\mathbf{Q},\tau)|^2 = (g_2(\mathbf{Q},\tau)-1)/A$, via the Siegert relation $g_2(\mathbf{Q},\tau) =1+A|F(\mathbf{Q},\tau)|^2 = 1+A|\exp[-(\tau/\tau_{r})^{\beta}]|^2$, which is displayed in Fig.~4\textbf{b,c} in the Main. Nonlinear least-squares using the
Levenberg--Marquardt algorithm as implemented in \texttt{scipy.optimize.curve\_fit}~\cite{virtanen_scipy_2020} was used to fit the exponential decays. 

\begin{table}[ht]
\centering
\caption{Details and numerical results from the KWW fits to the measured $g_2(\mathbf{Q},\tau)$ autocorrelation functions from the central partition for \bmNPS at all temperatures. Only the leading decay of the measured autocorrelation function was fit, with $\tau_{\mathrm{max}}$ indicating the maximum value of the fit range.}
\label{tab:bm_nps_fits}
\begin{tabular}{lcccc}
\toprule
$T$ $(K)$ & Fit $\tau_{\mathrm{max}}$ $(s)$ & $A$ & $\tau_r$ $(s)$ & $\beta$ \\
\midrule
298 & 44 & $0.768\pm0.042$ & $43.1\pm2.9$ & $2.037\pm0.098$ \\[8pt]
305 & 176 & $0.755\pm0.039$ & $217.0\pm13.2$ & $2.008\pm0.056$ \\[8pt]
325 & 176 & $0.640\pm0.019$ & $252.3\pm27.2$ & $2.039\pm0.100$ \\[8pt]
345 & 224 & $0.632\pm0.009$ & $189.0\pm13.7$ & $1.734\pm0.021$\\[8pt]
365 & 128 & $0.789\pm0.041$ & $144.9\pm10.5$ & $1.501\pm0.086$ \\[8pt]
385 & 864 & $0.687\pm0.043$ & $893.3\pm35.2$ & $1.145\pm0.327$ \\[8pt]
405 & 768 & $0.790\pm0.028$ & $1360.7\pm212.8$ & $1.122\pm0.100$   \\[8pt]
350-Cooling & 384 & $0.973\pm0.023$ & $(1.019\pm0.711)\times10^4$ & $0.983\pm0.087$ \\
\bottomrule
\end{tabular}
\end{table}

Table~\ref{tab:fit_ht} below displays the fit result to the single, central partition used for the highly-ordered \htNPS Bragg peak. The same fit protocol and error estimation was used for \htNPS, with the exception of a modified Siegert relation using a variable baseline, $g_2(\mathbf{Q},\tau) =c+A|F(\mathbf{Q},\tau)|^2 = c+A|\exp[-(\tau/\tau_{r})^{\beta}]|^2$. A variable baseline was necessary due to the compact speckle morphology of the \htNPS 211 peak leading to higher contrast influenced by the strong intensity variation of the sharp Bragg peak. 

\begin{table}[ht]
\centering
\caption{Details and numerical results from the KWW fits to the measured $g_2(\mathbf{Q},\tau)$ autocorrelation functions for the heat-treated \htNPS, displayed as the normalized Intermediate Scattering Function (ISF) in Main.}
\label{tab:fit_ht}
\begin{tabular}{lccccc}
\toprule
$T$ $(K)$ & Fit $\tau_{\mathrm{max}}$ $(s)$ & $A$ & $\tau_r$ $(s)$ & $\beta$ & $c$  \\
\midrule

298 & 384 & $2.175\pm0.506$ & $(2.190\pm0.583)\times10^3$ & $0.664\pm0.146$ & $0.041\pm0.304$\\
\bottomrule
\end{tabular}
\end{table}

\section{Recurrence interval from residuals of the leading decay}\label{snote:recurrence}

Oscillatory structure in correlation functions is often treated with a heterodyne mixing term, in which a static reference field beats against the fluctuating signal and the intermediate scattering function enters linearly rather than squared \cite{song_one-dimensional_2019,busch_dynamics_2008}. Such experiments are typically performed in small-angle geometries with a reference scatterer deliberately introduced into the beam path. No equivalent static reference field is present in the Bragg geometry used here, where the scattered amplitude originates entirely from the single illuminated primary particle, so the heterodyne form is not physically motivated in this measurement. Neither are the oscillations homodyne in form, where externally applied shear leads to observed oscillatory behavior. All of our measured dynamics are believed intrinsic and are not well described by homo- or heterodyne mixing.

We nonetheless tested a heterodyne form against \gTwoFull extracted from the diagonals of \TTCFull, as described in Methods. Averaging windows centered at $t_{age}=$~\qty{900}{\second} of varying sizes spanning the entire measurement time of $\pm$\qty{900}{\second} to small, high-fidelity segments of size $\pm$\qty{10}{\second} were investigated. None described the data better than the single KWW exponential \KWW, and in particular failed to fit the contrast and baseline. We therefore fit the leading relaxation with the single KWW form throughout and treat the recorrelations separately. This choice is necessary for consistency across the temperature series, since the additional parameters of a mixing model are undefined at temperatures where the recorrelations have degraded and in \htNPS where they are absent, whereas \taur and \shape are well defined at every measured temperature.


\begin{figure}[H]
	\centering
	\includegraphics[width=\linewidth]{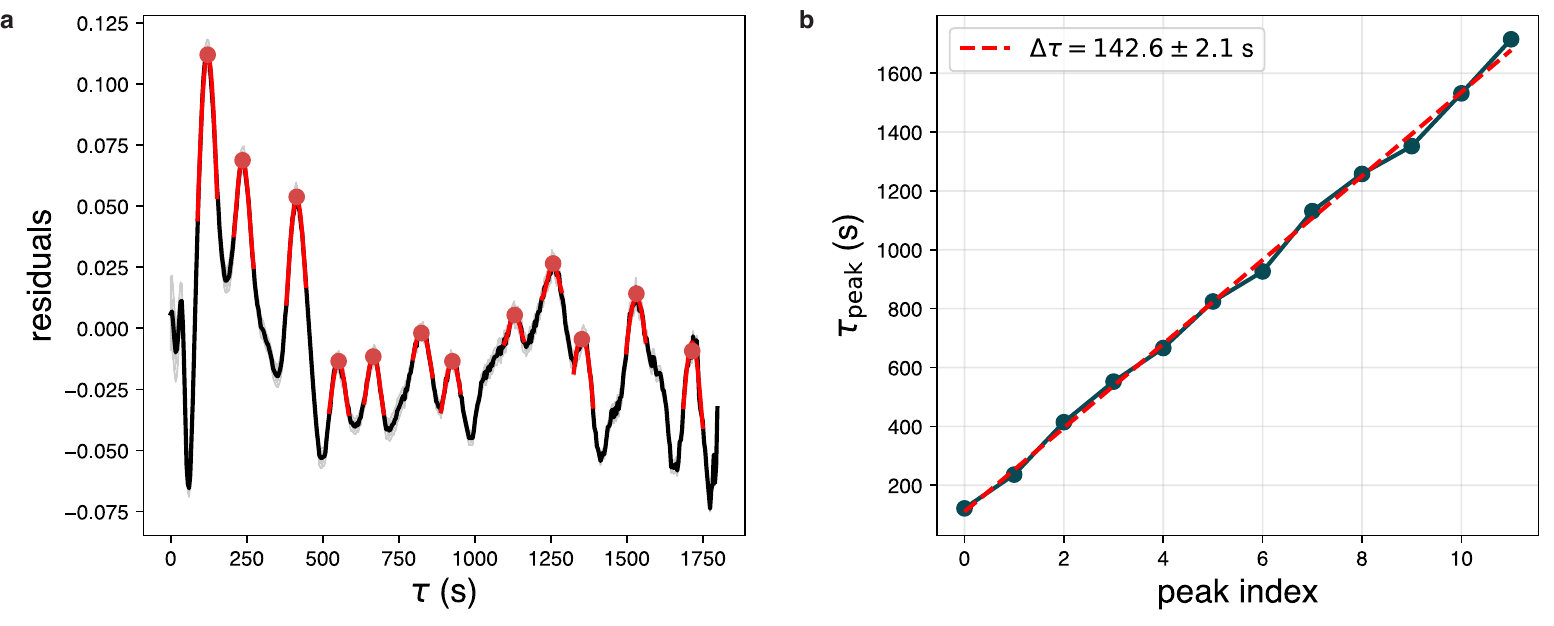}
	\caption{\textbf{Recurrence Interval in the Ball-Milled Sample.} \textbf{a} Residual of \ISFIP at \qty{298}{\kelvin} after subtraction of the fitted leading compressed exponential, with detected maxima marked. \textbf{b} Delay time of each maximum as a function of peak index. The dashed line is a weighted linear fit with slope $\Delta\tau = \qty{142.6}{\second}$.}
	\label{sfig:recurrence}
\end{figure}

To characterize the recurrences following the leading decay, \gTwoFull at \SI{298}{\kelvin} was extracted from the diagonals of \TTCFull rather than from the logarithmically binned multi-tau correlation, so that the long-lag structure is sampled uniformly in $\tau$. The residual is the measured intermediate scattering function subtracted by its fit, $\text{residuals}(\tau)=|F(\boldsymbol{Q},\tau)|^2_{\mathrm{cut}}-|F(\boldsymbol{Q},\tau)|^2_{\mathrm{fit}}$. Peaks are identified only for $40\,\mathrm{s}<\tau<1780\,\mathrm{s}$. Within this window, local maxima are located with a prominence threshold and a minimum separation. The prominence threshold is $15\%$ of the standard deviation of the residual, and is never smaller than $10^{-4}$. Neighboring maxima must be separated by at least five samples. Each maximum is then refined by fitting a Gaussian over a window of eight samples on either side,
\begin{equation}
A\exp\!\left[-\frac{1}{2}\left(\frac{\tau-\tau_0}{\sigma}\right)^2\right]+c.
\end{equation}
The initial amplitude is the peak height above the local median, the initial center is the detected delay, and the initial width is the larger of $5\%$ of that delay and one sixth of the fitting window. The amplitude is constrained to be positive and the center is constrained to lie inside the window. If the fit does not converge, the unrefined detected delay is retained. Accepted peaks are numbered in order, $n=0,1,2,\ldots$. Their delays are fit to a straight line,
\begin{equation}
\tau_n=\tau_0+n\Delta\tau,
\end{equation}
 where the slope $\Delta\tau$ is the quasi-period and the uncertainty on the spacing $\Delta\tau$ is the standard statistical error of the slope.

\section{Avalanche threshold and event detection}\label{snote:avalanche}

\begin{figure}[H]
	\centering
	\includegraphics[width=\linewidth]{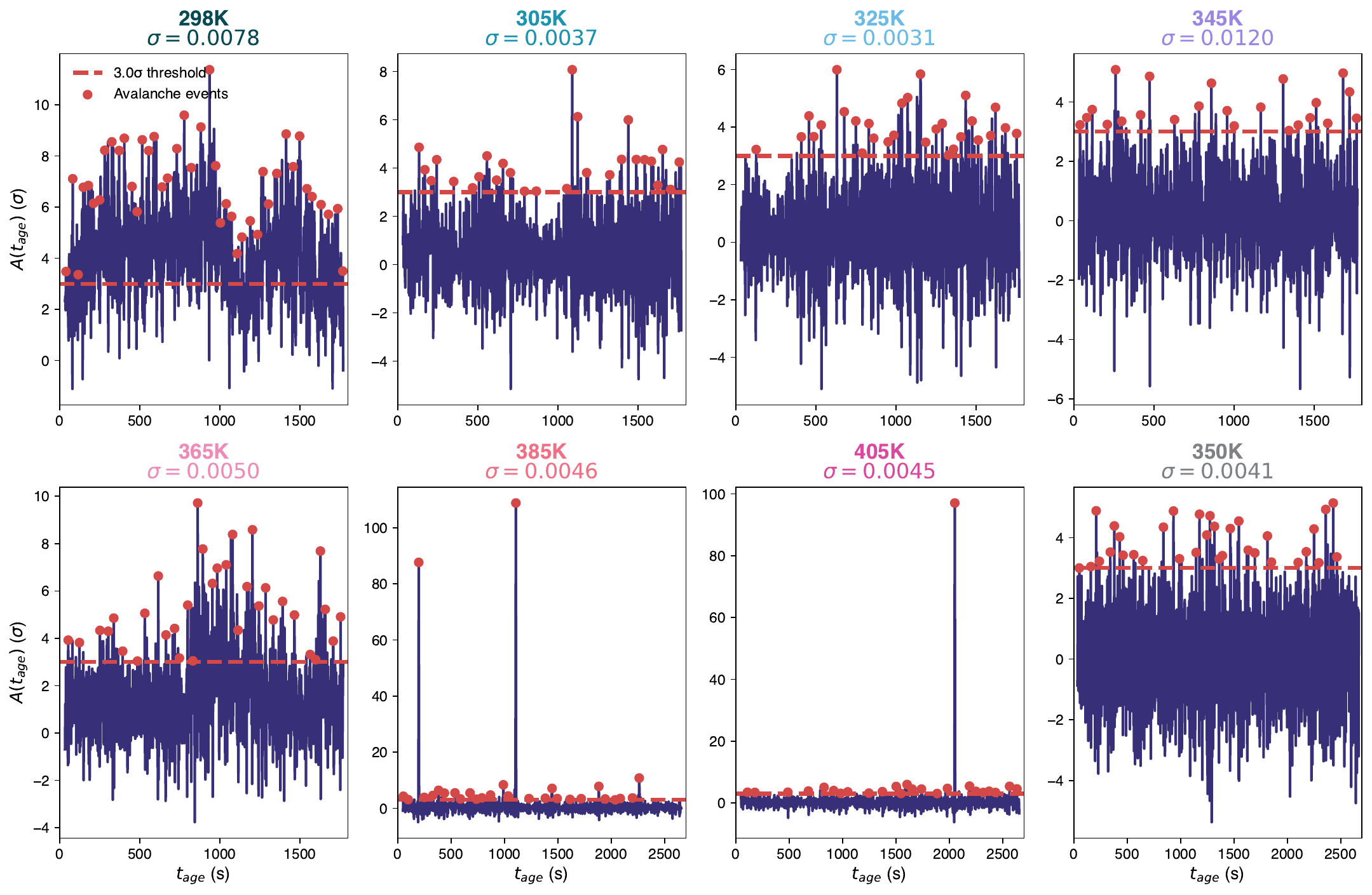}
	\caption{\textbf{Avalanche Detection for All Temperatures.} Full temperature series showing detected avalanche events from the two-time correlation matrices \TTCFull using the techniques detailed in Methods (Main) with an averaging window size of $m=$~\qty{15}{s}, for consistency with \cite{sanborn_direct_2011}. The standard deviation of the lower 50th percentile of fluctuations ($\sigma$), considered the noise floor, is provided in the title for each temperature. The detected amplitude variation is normalized by $\sigma$ and anything over 3$\sigma$ is considered a significant event (red dashed line). A minimum separation of \qty{15}{\second}, equal to $m$, is used to distinguish unique events.  }
	\label{sfig:avalanches}
\end{figure}

For completeness, all avalanche detection spectrograms $A(\tage)$ $(\sigma)$ are provided here, the calculation of which is described in Methods in Main. These figures serve to show that the frequent smaller events less than approximately 10, in units of the noise floor $\sigma$, dominate the avalanche activity across all temperatures where large system spanning events are much less likely.

\section{Sample stability during acquisition}\label{snote:stability}

\begin{figure}[ht]
	\centering
	\includegraphics[width=\linewidth]{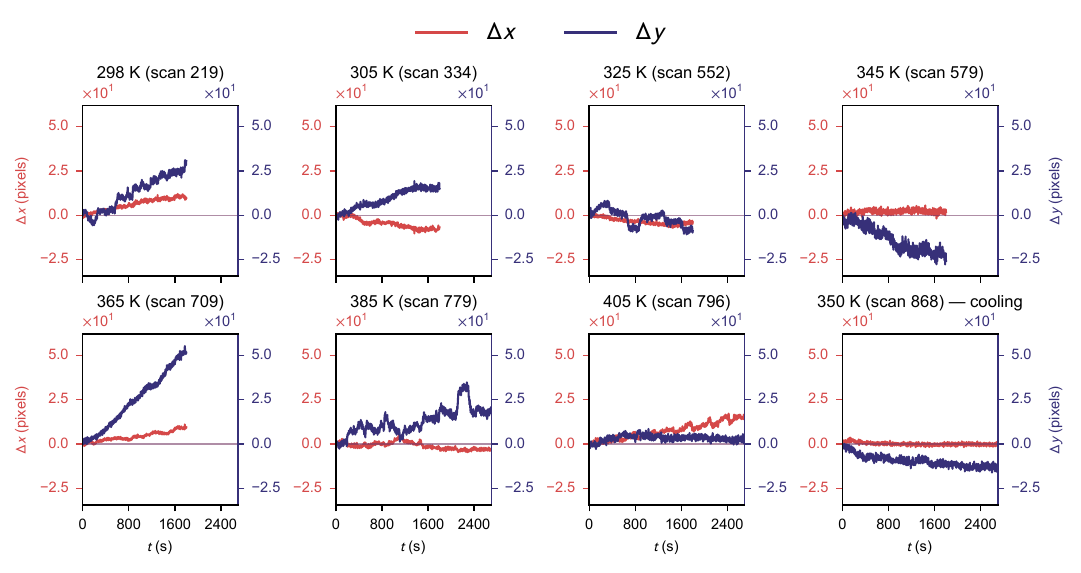}
	\caption{\textbf{Center of Mass Evolution.} Intensity-weighted centroid position of the 211 reflection as a function time for all temperatures. Horizontal lines indicate the initial positions $\Delta Q_{x} = \Delta Q_{y} = 0$.}
	\label{sfig:com}
\end{figure}

While the sample environment, beamline, and stage for this series of experiments is stable based on our measurement of static references, including the room-temperature \htNPS measurement and the final measurement in the annealing series for \bmNPS at \qty{350}{\kelvin}, additional checks were performed to ensure the observed oscillatory behavior is intrinsic to the measured mesoscale dynamics in \bmNPS. In Fig.~\ref{sfig:com}, we have the change in centroid center of mass (COM) of the scattered intensity in detector pixel-space plotted as a function of measurement time. The COM is calculated using the intensity weighted average,
\begin{equation}\label{seq:com}
    \langle x\rangle_{I}(t)=\frac{\sum_p I_{p}(t)x_{p}}{\sum_p I_{p}(t)},
\end{equation}
and likewise for $\langle y\rangle_{I}(t)$. Rather than plotting the absolute COM evolution in $(x,y)$, we show the change in the COM defined as $\Delta x(t) = \langle x\rangle_{I}(t) - \langle x\rangle_{I}(0)$, and likewise for $\Delta y(t)$. This diagnostic proves invaluable when informed by the relaxation times from the KWW fit results in Table~\ref{tab:bm_nps_fits}, the quasi-periodic recurrence events extracted in Fig.~\ref{sfig:recurrence}, and the characteristic speckle size for \bmNPS calculated in Section~\ref{snote:particle_size}. 

As an example we focus on \bmNPS for \qty{298}{\kelvin}, which possess the fastest dynamics measured and the most well-defined quasi-periodicity, and therefore is more susceptible to drift or reorientaion contaminating the measurement. From Fig.~\ref{sfig:com}, we see that over the course of the \qty{1800}{\second} measurement, a COM change for the 211 \bmNPS peak of $\Delta x \approx \num{12}$ pixels and $\Delta y \approx \num{30}$ pixel is observed, corresponding to a change in \Q-space of $\Delta \Q_{x} \approx -1.2\times10^{-3}$\unit{\per\angstrom} and $\Delta \Q_{y} \approx -3\times10^{-3}$\unit{\per\angstrom}, respectively, well within the total speckle extent of $\Delta Q \approx 2\pi/\corrLength \approx 2\pi/\qty{34.2}{\nano\meter} \approx 1.84\times10^{-2}$\unit{\per\angstrom}. This implies that the structural symmetry has not changed over the course of the measurement and the orientation remains within the initial Bragg peak profile. Furthermore, recall that the characteristic speckle size for \bmNPS is approximately $\Delta l_{px} \approx 6$ pixels. Observing a monotonic, roughly linear drift in the COM at \qty{298}{\kelvin}, the rate of change of the COM is approximately $\Delta x/\Delta t \approx 12/1800\unit{\second} \approx 6.67\times 10^{-3}$ pixels~\unit{\per\second} and $\Delta y/\Delta t \approx 30/1800\unit{\second} \approx 1.67\times 10^{-2}$ pixels~\unit{\per\second}. Considering the faster rate, this means that the COM traverses a distance the size of a single speckle in approximately $t_{speckle} = \Delta l_{px}/\Delta y/\Delta t = 6~\text{pixels}/1.67\times 10^{-2}~\text{pixels}~\unit{\per\second}\approx \qty{360}{\second}$, or \qty{6}{\minute}. Thus, $t_{speckle}$ is $\sim\num{8.35}$ times longer than the measured relaxation time \taur = \qty{43.1}{\second} for \bmNPS at \qty{298}{\kelvin} and $\sim\num{2.52}$ times longer than the quasi-period $\Delta \tau$ = \qty{142.6}{\second} in Fig.~\ref{sfig:recurrence}. Therefore, we conclude that the measured dynamics including \taur and the quasi-period $\Delta \tau$ cannot be attributed to changes in the COM alone, which itself may very well be intrinsic, and that this argument holds for all other temperatures where the relaxation and periodicity are notably longer and the changes in the COM smaller as the initially strained configuration is relaxed. 

\begin{figure}[tbp]
	\centering
	\includegraphics[width=\linewidth]{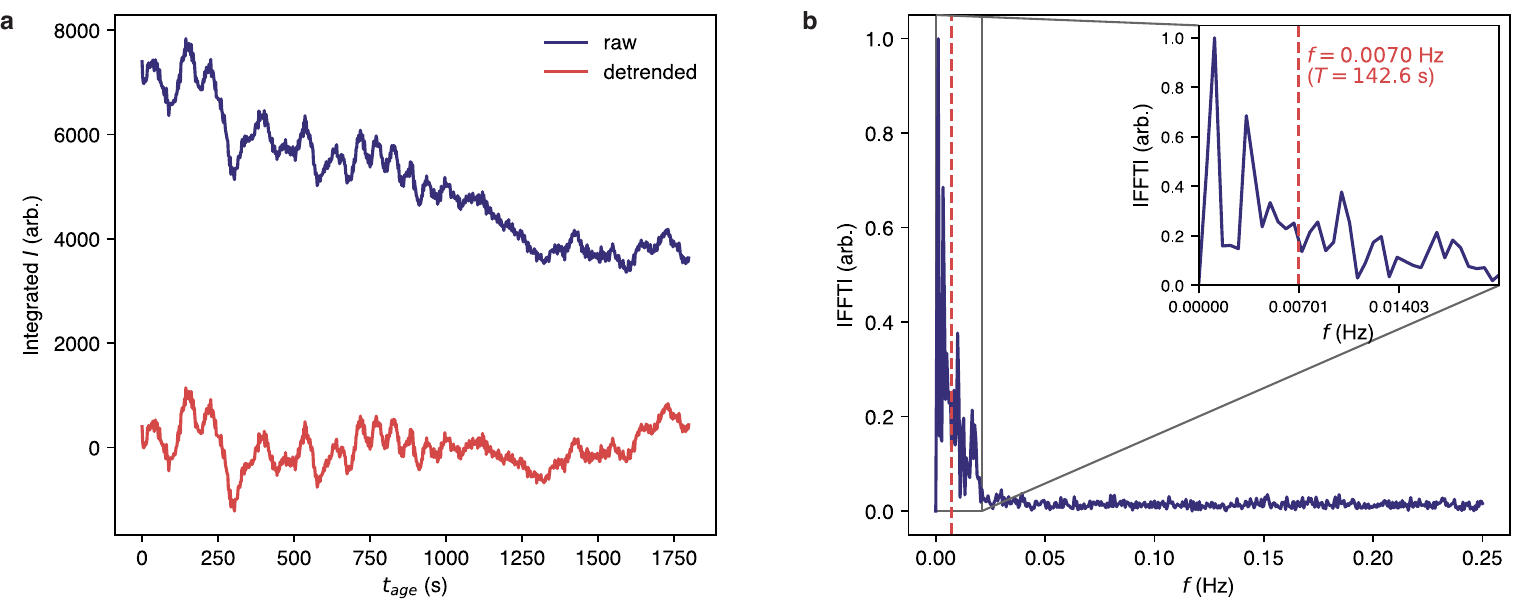}
	\caption{\textbf{Integrated Intensity and Power Spectra.} \textbf{a} Integrated intensity of all pixels for all frames within the central partition for \bmNPS at \qty{298}{\kelvin}. Both the raw and detrended intenities are shown. \textbf{b} Normalized power spectrum of the detrended integrated intensity in \textbf{a}, showing no significant peak in the power spectrum at the \SI{142.6}{\second} recurrence interval.}
	\label{sfig:int_fft}
\end{figure}

In addition to changes in the COM of the measured speckle pattern, inspecting the raw integrated intensity versus measurement time for the central partition over which \gTwoFull and \TTCFull are calculated is also informative. Again, considering only the stability of \bmNPS at \qty{298}{\kelvin} as the measurement of most concern, we see the raw integrated intensity in Fig.~\ref{sfig:int_fft}\textbf{a} decreases by approximately \qty{50}{\percent} over the course of the \qty{1800}{\second} measurement. While this could be due to factors mentioned earlier such a beam damage, drift of the particle or beam, or rotation causing misalignment, usually only order of magnitude changes are a concern. The relatively slow decay of only $\sim$\num{2.2}~\text{counts}~\unit{\per\second} and the monotonic decay makes the loss in intensity manageable as both \gTwoFull and \TTCFull are normalized against changes in intensity as detailed in Methods. 

Often oscillatory behavior observed in \TTCFull matrices can be directly attributed to oscillations in the integrated intensity. To preclude this possibility, we calculated the power spectrum as the absolute value of the Fast Fourier Transform of the detrended integrated intensity to observe if any peaks are present at the quasi-period of $\Delta \tau = \qty{142.6}{\second}$. The intensity was linearly detrended using \texttt{scipy.signal.detrend} to remove the DC background. The inset in Fig.~\ref{sfig:int_fft}\textbf{b} shows no significant power is present at $1/\Delta \tau = \qty{0.007}{\hertz}$ indicating the oscillations observed in \TTCFull at \qty{298}{\kelvin} are not driven by oscillations in intensity. 

\bibliography{sn_bibliography}